\documentstyle[12pt]{article}
\newskip\humongous \humongous=0pt plus 1000pt minus 1000pt
\def\caja{\mathsurround=0pt}

\newif\ifdtup
\def\panorama{\global\dtuptrue \openup1\jot \caja
	\everycr{\noalign{\ifdtup \global\dtupfalse
	\vskip-\lineskiplimit \vskip\normallineskiplimit
	\else \penalty\interdisplaylinepenalty \fi}}}
\def\eqalignno#1{\panorama \tabskip=\humongous
	\halign to\displaywidth{\hfil$\displaystyle{##}$
	\tabskip=0pt&$\displaystyle{{}##}$\hfil
	\tabskip=\humongous&\llap{$##$}\tabskip=0pt
	\crcr#1\crcr}}

\def\eg{\hbox{\it e.g.}}

\let\vev\VEV

\def\abs#1{\left| #1\right|}

\def\ltap{\raisebox{-.4ex}{\rlap{$\sim$}} \raisebox{.4ex}{$<$}}

\begin{document}
\begin{titlepage}
\today          \hfill 
\begin{center}
\hfill    LBL-38110 \\
          \hfill    UCB-PTH-96/01 \\

\vskip .5in

{\large \bf The Heavy Top Quark And Supersymmetry}\footnote{Lectures given at
the 1995
SLAC Summer Institute, July 10-21st, 1995.}
\footnote{This work was supported in part by the Director, Office of 
Energy Research, Office of High Energy and Nuclear Physics, Division of 
High Energy Physics of the U.S. Department of Energy under Contract 
DE-AC03-76SF00098 and in part by the National Science Foundation under 
grant PHY-95-14797.}

\vskip .5in
Lawrence J. Hall

{\em Theoretical Physics Group, Lawrence Berkeley Laboratory\\
  and Department of Physics, University of California, Berkeley, CA 94720\\
     E-mail: hall\_lj@lbl.gov}
\end{center}

\vskip .5in

\begin{abstract}
Three aspects of supersymmetric theories are discussed: electroweak symmetry
breaking, the issues of flavor, and gauge unification. The heavy top quark
plays an important, sometimes dominant, role in each case.
Additional symmetries lead to extensions of the standard model which can provide
an understanding for many of the outstanding problems of particle physics. A
broken supersymmetric extension of spacetime allows electroweak symmetry
breaking to follow from the dynamics of the heavy top quark; an
extension of isospin provides a constrained framework for understanding the
pattern of quark and lepton masses; and a grand unified extension of the
standard model gauge group provides an elegant understanding of the gauge
quantum numbers of the components of a generation. Experimental signatures for
each of these additional symmetries are discussed.
\end{abstract}
\end{titlepage}
\renewcommand{\thepage}{\roman{page}}
\setcounter{page}{2}
\mbox{ }

\vskip 1in

\begin{center}
{\bf Disclaimer}
\end{center}

\vskip .2in

\begin{scriptsize}
\begin{quotation}
This document was prepared as an account of work sponsored by the United
States Government. While this document is believed to contain correct 
 information, neither the United States Government nor any agency
thereof, nor The Regents of the University of California, nor any of their
employees, makes any warranty, express or implied, or assumes any legal
liability or responsibility for the accuracy, completeness, or usefulness
of any information, apparatus, product, or process disclosed, or represents
that its use would not infringe privately owned rights.  Reference herein
to any specific commercial products process, or service by its trade name,
trademark, manufacturer, or otherwise, does not necessarily constitute or
imply its endorsement, recommendation, or favoring by the United States
Government or any agency thereof, or The Regents of the University of
California.  The views and opinions of authors expressed herein do not
necessarily state or reflect those of the United States Government or any
agency thereof, or The Regents of the University of California.
\end{quotation}
\end{scriptsize}

\vskip 2in

\begin{center}
\begin{small}
{\it Lawrence Berkeley Laboratory is an equal opportunity employer.}
\end{small}
\end{center}

\newpage
\renewcommand{\thepage}{\arabic{page}}
\setcounter{page}{1}

\centerline{\bf Table of Contents}

\vskip .25in

\centerline{\bf (I) Symmetries \& Symmetry Breaking}

\vskip .25in

\begin{enumerate}
\item Symmetries
\item  Flavor Symmetries
\item  The major problems of the high energy frontier
\item Supersymmetry
\item Summary
\end{enumerate}

\vskip .25in

\centerline{\bf (II) $SU(2) \times U(1)$ Breaking and the Weak Scale}

\begin{enumerate}
\item A symmetry description
\item Matter v. Higgs
\item A heavy top quark effect
\end{enumerate}

\vskip .25in

\centerline{\bf (III) Flavor In Supersymmetric Theories}

\vskip .25in

\begin{enumerate}
\item The fermion mass and flavor-changing problems
\item Approximate flavor symmetries
\item The flavor-changing constraints
\item The maximal flavor symmetry
\item A brief introduction to the literature
\item The minimal $U(2)$ theory of flavor
\item The suppression of baryon and lepton number violation
\item Conclusions
\item Appendix A
\end{enumerate}

\vskip .25in

\centerline{\bf IV. Supersymmetric Grand Unification}

\vskip .25in

\begin{enumerate}
\item Introduction
\item Flavor signals compared
\item Flavor-changing and CP violating signals
\item The top quark origin of new flavor and CP violation
\item Summary.
\end{enumerate}

\vskip .25in

\centerline{\bf V. The High Energy Frontier}

\vskip 1in

{\large \bf I. Symmetries and Symmetry Breaking}

\noindent {\bf I.1 Symmetries}

Much progress in particle physics has been made possible by understanding
phenomena in terms of symmetries, which can be divided into four types:
global or local action in spacetime or in an internal space.
A symmetry of any of these types can be further classified as exact or broken,
according to whether any breaking has been measured in experiments,
as illustrated by well-known examples in Table 1.
In these lectures I discuss three of the four symmetry types, leaving out the
gauging of spacetime symmetries which is expected to occur at the Planck scale.

An interesting feature of Table 1 is that of the six entries, only five have
been discovered in nature: there is no experimental evidence for a broken,
global symmetry of spacetime, hence the blank entry.

\newpage
\begin{center}
{\bf Table 1 Symmetries}
\vskip 9pt
\begin{tabular}{|c|c|c|}
\hline
&{\bf{EXACT}}&{\bf{BROKEN}}\cr
\hline
Local&$SU(3)_{QCD}$&$SU(2)\times U(1)_Y$\cr
Internal&$U(1)_{EM}$&\cr
\hline
Global&Baryon number: $B$&Isospin: $SU(2)_I$\cr
Internal&Individual lepton numbers: $L_i$&\cr
\hline
Global&Displacements: $P$&\cr
Spacetime& Angular momentum: J&\cr
& Lorentz boosts: K&\cr
\hline
\end{tabular}
\end{center}

\medskip

\noindent {\bf I.2 Flavor Symmetries }

With one exception, the entries of Table 1 provide a complete list of what has
been discovered experimentally for these categories, ignoring the discrete
spacetime symmetries such as parity.
The exception is provided by global internal symmetries. 
Including color and weak degrees of freedom, 45 species of quarks and leptons 
have been 
found; experiments have therefore uncovered a $U(45)$ global internal, or 
flavor, symmetry,
which is broken to $B\times L_i$ by the known gauge interactions and particle 
masses.
The existence and masses of these 45 states, together with the way the known
gauge forces act on them, is the flavor puzzle of particle physics. It is 
instructive to 
consider separately the breaking of $U(45)$ by gauge interactions and by masses.
The known gauge interactions divide the 45 states into 3 identical periods, 
or generations, 
each of which contains five multiplets transforming irreducibly under the 
gauge group: $q,u,d,l,e$, as shown in Table 2.
I have chosen to write each fermion as a left handed spinor of the Lorentz 
groups, so that $u,d$, and $e$ are left-handed anti-quarks and anti-leptons.
In Table 2 the number of states 
for each of the five representations is shown in parenthesis,
the total being 15 for each of the three generations.

\newpage

\begin{center}
{\bf{ Table 2  \  The Aperiod Table.}}
\vskip 9pt
\begin{tabular}{|c|c|c|r|}
\hline
&$SU(3)$&$SU(2)$&$U(1)_Y$\cr
\hline
$q(6)$&3&2&${1\over 6}$\cr
\hline
$u(3)$&$\overline{3}$&---&$-{2\over 3}$\cr
\hline
$d(3)$&$\overline{3}$&---&${1\over 3}$\cr
\hline
$l(2)$&---&2&$-{1 \over 2}$\cr
\hline
$e(1)$&---&---&1\cr
\hline
\end{tabular}
\end{center}

\noindent The known gauge interactions distinguish between the
15 states of a generation, but do not distinguish
between the three generations; they break 
the flavor symmetry group from $U(45)$ to $U(3)^5$, with one $U(3)$ 
factor acting in generation space on each of the five multiplets $q,u,d,l,e$.

This $U(3)^5$ symmetry is broken in hierarchical stages by the quark and lepton 
mass matrices.
For example, the up quark matrix provides an explicit breaking
of $U(3)_q \times U(3)_u$ transforming as a (3,3).
The largest entry in the matrix is clearly the top
quark mass, which strongly breaks this group to $U(2)_q \times U(2)_u \times
U(1)_{q_3 - u_3}$.
The fermion mass problem, which is part of the flavor puzzle, is the question 
of why the quark and lepton mass matrices 
break $U(3)^5$ in the hierarchical fashion measured by experiment.
Since we are dealing with matrices,
a solution of this problem would provide an 
understanding of both quark and lepton masses and the
Kobayashi-Maskawa mixing matrix.
All questions about the quark and lepton masses and mixings can be rephrased
in terms of $U(3)^5$ breaking.
For example why is $m_t \gg  m_b$ becomes: why is the breaking
$U(3)_u \to U(2)_u$ stronger than that of $U(3)_d \to U(2)_d$?
In the context of the standard model this rephrasing does not seem very
important; however, in the context of supersymmetry it is of great importance.

\medskip

\noindent {\bf I.3 The major problems of the high energy frontier}

\medskip

All physicists should spend a great deal of time debating and deciding what 
are the most important issues in their subfield.
At the high energy frontier, I think the four most import puzzles are

\begin{enumerate}
\item What breaks $SU(2) \times U(1)$?

The weak interactions appear weak, and are short range, because they, alone 
among the known forces, are generated from a symmetry
group which is broken.
Perturbative gauge forces do not break themselves: new interactions are 
required to break them.
Such a fifth force must exist and be accessible to experiments
designed to probe the weak scale.
It is guaranteed to be exciting: it has a dynamics
which is different from any of the known forces, and it should shed light on the
fundamental question of what sets the mass scale of weak symmetry breaking.
I will call this mass scale $M_Z$, even though the weak symmetry breaking 
mechanism
of the fifth force is responsible for the dominant contribution to the mass of 
all of the known massive elementary particles.

\item What breaks the $U(3)^5$ flavor symmetry?

We know that this flavor symmetry is broken at least to $B\times L_i$ because 
of the observed quark and lepton masses and the Kobayashi-Maskawa mixing matrix.
However, such masses and mixings cannot simply be inserted into the 
theory because they break
$SU(2)$; they must originate from some new interactions which break $U(3)^5$.
In the standard model these new interactions are the Yukawa couplings of the 
Higgs boson, but there are other possibilities.
We might call these $U(3)^5$ breaking interactions the ``sixth force''.
I think that future experiments will uncover this force also, at least the 
pieces of it which are strong and are responsible for the large top quark mass.
Whatever the description of $U(3)^5$ breaking at the weak scale, there is
still the puzzle as to why $U(3)^5$ is hierarchically broken.
I think that physics at the weak scale could shed light on some aspects of this;
but this is much more uncertain.
It is likely that some, and perhaps all, of the understanding of flavor 
physics occurs at some very much higher energy scale.
Nevertheless, at the very minimum, experiments must be done which uncover the
weak scale description of $U(3)^5$ breaking, ie the sixth force.
I find a sense of excitement building up in our field as experiments enter the
domain where signals of the fifth and sixth forces will be discovered.

\item  Why are the symmetries and fundamental constants of nature what they 
are?

The most basic properties of nature can be summarized in terms of a set of
gauge, flavor and spacetime symmetries, and a set of fundamental parameters,
such as the gauge couplings and the quark and leptons masses. The next question 
is embarrassingly obvious: why these symmetries and why these values of the 
parameters?
The anthropic argument, that without them we could not exist to make the 
observations, is fraught with problems; it seems to me
better to look boldly for a true theory. 
A complete answer to these questions requires going beyond 
four dimensional, point particle quantum  field theory, and at the moment 
superstring theory provides the unique such direction.
However, string theory is very ambitious, and, despite exciting developments,
the time scale for making definitive connections to physics is completely
unknown.  
The central thesis of these lectures is that
we may already have the basic tools required to make considerable progress
in furthering our understanding of nature. The familiar tools of unified gauge
symmetries, flavor symmetries, the properties of supersymmetry and the 
renormalization group can carry us very far, and can be tested by experiment.
The gauge group $SO(10)$ explains the quantum numbers of Table 2. 
If the 15 known states of a generation, together with a right handed neutrino,
are placed in the 16 dimensional spinor representation of $SO(10)$, then every
entry of Table 2 follows from the simple group theoretic embedding of $SU(3)
\times SU(2) \times U(1)$ into $SO(10)$. This is an extraordinary achievement.
The vertical unification of a generation also reduces the flavor symmetry 
group from $U(3)^5$ to $U(3)$, which is much more constraining.
Such grand unified theories can reduce the number of free 
parameters on which all of low energy physics depends.
Several supersymmetric theories based on the flavor group $U(3)$, or on one of
its subgroups, have been developed recently, and make many predictions for the
flavor changing interactions of the superpartners. Such grand unified theories
of flavor are not
the ultimate theory, but they can explain a great deal very simply.
For grand unified and flavor symmetries the real question is: 
how can they be subjected to experimental tests? I will begin the 
answer to this question in these lectures.

\item How is a quantum theory of gravity to be constructed?

Superstring theory provides the only known direction for progress.
\end{enumerate}

\noindent {\bf I.4 Supersymmetry}

\medskip

The current interest in supersymmetry is largely because it offers 
interesting new directions for attacking each of the above problems.
In summary these new directions are 

\begin{enumerate}
\item Supersymmetry is the only symmetry which can give rise to a light,
elementary Higgs boson for electroweak symmetry breaking.
The puzzle of the scale of weak interactions is replaced with the puzzle of the 
origin of the scale of supersymmetry breaking.

\item The hierarchical breaking of $U(3)^5$ governs not only the form of the 
Yukawa interactions of the Higgs, but also the squark and slepton mass matrices.
Since the latter are severely constrained by flavor changing phenomenology,
severe restrictions are placed on the group theoretic structure of the pattern 
of 
$U(3)^5$ breaking. In addition, supersymmetry allows for the possibility that 
above the weak scale some of the $U(3)^5$ breaking which generates the quark 
and lepton masses arises from the scalar mass matrices rather than from the 
Higgs Yukawa interactions.

\item Supersymmetric grand unification provides a successful prediction, at 
the percent level, of the weak mixing angle.
Although less significant, $m_b/m_t$ and $m_t$ can also be successfully 
predicted in supersymmetric unified models.
With further simplifying assumptions, such as the nature and breaking of the 
flavor group, other predictions can also be obtained.
\item A supersymmetric string theory offers the prospect of a 
quantum theory of gravity, unified with the other forces.
\end{enumerate}

In these lectures, I will elaborate on the first three of the above: 
$SU(2) \times U(1)$ breaking, flavor symmetry breaking and
supersymmetric grand unification, in chapters II, III, and IV respectively.

There are many excellent books and review articles on supersymmetry \cite{WB}, 
the supersymmetric extension of the standard model \cite{HKane} 
and supersymmetric grand unification.
The aim of the present lectures is not to refine or update these works,
but to explain why I think the study of supersymmetry is interesting, why the
direct search for superpartners is of crucial importance, and what may be learnt
from a variety of other measurements.
Nevertheless, it may be useful to say a few words about supersymmetry and 
the supersymmetric extension of the standard model.

Supersymmetry is an extension of the Poincare group of spacetime 
transformations.
Spinorial generators, $Q$ and $\overline{Q}$, are added to the usual generators 
$p, J$ and $K$ of translations, rotations and boosts.
The only non-trivial extension of the Poincare algebra involving $Q$ or
$\overline{Q}$ is the anticommutation $\{Q, \overline{Q}\}=p$.
Consider the evolution of our understanding of the spacetime properties of the 
electron. When discovered, nearly a century ago
\footnote{ I expect we will have celebrations in 1997 for the centenary of the
discovery of the first particle which, as far as we know today, is 
elementary.},
by J.J. Thompson, it was conceived as a negatively
charged particle with just two properties:
its mass and electric charge.
We view the charge as a consequence of the behavior with respect to the
electromagnetic $U(1)$ charge generator, and the mass as a consequence of the 
translation generator $p$.
The discoveries of Stern and Gerlach dictated that it should be given 
another attribute, 
intrinsic spin, which describe its properties with respect to the angular 
momentum generator, $J$.
The splitting of an atomic beam by an inhomogeneous magnetic field, which they
discovered in 1922, is caused by the doubling of the number of electron states
which follows from their non-trivial properties under the angular momentum
generator: $ e  \stackrel{J}{ \longrightarrow} (e^\uparrow, e^\downarrow)$. 
In the relativistic case, this description is inadequate.
The Lorentz boost generator $K$ requires a further doubling of the number of 
particle states; we call the resulting Lorentz-partners the antiparticles:
$e \stackrel{K}{ \longrightarrow} (e,\overline{e})$.
Their properties are dictated by Lorentz symmetry, having equal mass and 
opposite charge to the particles.

The extension of spacetime symmetries which results from the introduction 
of the supersymmetry generator, $Q$, causes a further doubling of the particles:
$ e \stackrel{Q}{\longrightarrow} (e, \widetilde{e})$;
while $\overline{e}$ is the Lorentz-partner of the electron, $\widetilde{e}$
is the supersymmetry-partner, or superpartner, of the electron.
It has properties which are determined by the supersymmetry algebra: the 
mass and charge are identical to that of the electron, but, because $Q$ 
is spinorial, it has intrinsic
spin which differs by 1/2 relative to the electron; it is a Lorentz scalar.
Many people laugh when they hear about supersymmetry and how it leads to the 
introduction of a new hypothetical particle for each of the observed particles.
However, it is just history repeating itself; perhaps physicists of old 
laughed at the prospect of antielectrons and antiprotons,
but the sniggering soon stopped.

The super-electron is not degenerate with the electron; supersymmetry, 
if it exists,
must be sufficiently broken that the selectron mass is larger than about 65 GeV.
The discovery of supersymmetry would be doubly exciting: not only would it 
herald an exciting new era of spectroscopy, but it would represent the
discovery of a completely new type of symmetry: a broken spacetime symmetry.
The empty box of Table 1 would be filled by $Q$; nature would have provided 
examples of all six varieties of symmetries. What could be more interesting?

\noindent {\bf I.5 Summary}

Three types of symmetries are shown in Table 1: local internal, global internal
and global spacetime, which I shall frequently call gauge, flavor and spacetime
symmetries, respectively.
Each of these types of symmetry may be broken at scales beneath the Planck 
scale $M_{Pl}$.
In these lectures I consider the breaking of a unified group,
$$
G_{unified} \; {\stackrel{M_G}{\longrightarrow}} \; SU(3) \times SU(2) 
\times U(1) \; {\stackrel{M_Z}{\longrightarrow}} \; SU(3) \times
U(1)\eqno(I.1)
$$
the breaking of the flavor symmetry group $G_f \subset U(3)^5$
$$
G_f \; {\stackrel{M_F}{\longrightarrow}} \; B\times L_i\eqno(I.2)
$$
and the breaking of supersymmetry
$$
(p, J, K , Q, \overline{Q}) \; \stackrel{M_S}{\longrightarrow} \; 
(p, J, K)\eqno(I.3)
$$
The mass scales represent the scales of the vacuum expectation values of
fields which break the symmetry.
There could be several stages of breaking of the unified gauge group, and there 
will almost certainly be several stages in the sequential
breaking of the flavor group, so $M_G$ and $M_F$ represent a set of scales.
Assuming that only one supersymmetry survives
beneath $M_{Pl}$, $M_S$ is unique.
In the limit that $M_S\to 0$, the superparticle and particle masses become 
degenerate; however in most schemes of supersymmetry
breaking, the mass scale $m_s$ of the superpartners of the known 
particles is not given by 
$M_S$.
For example in supergravity $m_s = M^2_S/M_{Pl}$ and in dynamical
supersymmetry breaking models $m_s = \alpha M^2_S/M_X$, where
$M_X$ is some other mass scale larger than $M_S$.
The scale $M_X$ or $M_{Pl}$ is known as the messenger scale, $M_{mess}$, 
it is the energy scale below which  the superpartners possess local 
supersymmetry breaking masses and interactions.

There is no guarantee that $M_F$ is less than $M_{Pl}$.
The physics of flavor may be understood only at the Planck scale. 
Indeed, of all the mass scales
introduced in this subsection, $M_F$ is perhaps the most uncertain.
If $M_F \approx M_{Pl}$, then $G_f$ breaking interactions must occur 
explicitly at the boundary at $M_{Pl}$, with small dimensionless coefficients.
An advantage to having $M_F$ beneath $M_{Pl}$ is that the small 
dimensionless fermion mass ratios can then appear as ratios of these scales.
In chapter III we will explore the case of $M_F < M_{Pl}$, which allows for an
understanding of at least some aspects of flavor beneath $M_{Pl}$.

\vskip 1in

{\large \bf  II. $SU(2) \times U(1)$ Breaking and the Weak Scale}

\noindent{\bf II.1 A Symmetry Description}

\medskip

In the standard model the $SU(2) \times U(1)$ electroweak symmetry is broken 
by introducing a 
Higgs sector to the theory, which involves an electroweak scalar doublet, $h$.
The mass squared parameter for this field, $m^2_h$, determines the order 
parameter
of the symmetry breaking: if it is negative the electroweak symmetry breaks,
while if it is positive all the elementary particles are massless.
The Higgs sector certainly provides an economical description of electroweak
symmetry breaking, but it is inadequate for two reasons.
There is no dynamical understanding of why symmetry breaking occurs; one simply
inserts it into the theory by hand by making $m^2_h$ negative.
Secondly, there is no symmetry understanding of the scale of the breaking,
which I refer to as the $Z$ mass, $M_Z$.

In physics we have learnt that that mass scales should be both described 
and understood in terms of symmetries. Great progress has been made in 
providing symmetry descriptions of phenomena, but understanding the origin 
of the symmetry behavior
at a deeper level often eludes us, as we illustrate with a few examples.

Why is the photon massless? The symmetry description is clear: electromagnetic 
gauge invariance is unbroken.
However, the deeper question is: {\it why} is it unbroken?
This brings us back to the breaking of $SU(2)\times U(1)$ electroweak symmetry.
Why is it accomplished by a single doublet, reducing the rank by one but 
not by two?

Why are the neutrinos massless?
A symmetry description is that nature possesses lepton number as an exact global
symmetry. At a deeper level, however, many questions arise: why are there no 
right-handed neutrinos,
why is lepton number exact. If the neutrinos do have small masses, why are the
lepton numbers such good approximate symmetries?
An interesting feature of supersymmetric theories is that the standard 
answers to these questions are inadequate, as discussed in II.2 and III.7.

Why do the quark and charged leptons have their observed masses?
Since the masses break the electroweak symmetry, they can be written as
$\lambda v$, where $v$ is the dimensionful order parameter of the
symmetry breaking and $\lambda$ is a dimensionless parameter,
different for each quark and lepton.
The overall scale of the masses is determined by $v$, while the
mass ratios are determined by ratios of $\lambda$ couplings.
Many of the $\lambda$ are small, which we describe in chapter III in terms 
of approximate 
flavor symmetries. But what is the origin for these symmetries and their 
breaking? Why are there three generations?
Why is the up quark so much lighter that the top quark: 
$\lambda_{up}/\lambda_{top} \approx 10^{-5}$?

What is the origin of the hadronic mass scale of the proton and neutron?
This scale is the scale at which the QCD coupling constant, $\alpha_s$,
becomes large and non-perturbative. 
It arises, through renormalization, as a dimensional transmutation of this
gauge coupling, and hence is described in terms of the QCD symmetry group, 
$SU(3)$.

These examples illustrate how we turn to symmetries for both a description 
and a deeper understanding of the phenomena.
This applies to all phenomena of particle physics, but here I stress 
the application to masses.

Now we can better appreciate the inadequacy of the standard model Higgs 
sector description of electroweak symmetry breaking.
What symmetry description or understanding does it proscribe for the order 
parameter $v$ which determines $M_Z$ and the fermion masses? {\it None.}
The crucial point is that it does not even provide a symmetry description 
for the scale $v$, let alone any deep understanding.
Because the standard model Higgs sector is so economical, and because the 
standard model provides
an accurate description of so much data, many have concluded that the standard
model will be the final story - there will be no physics beyond the standard 
model. I strongly disagree with this viewpoint.
First there is not a shred of evidence for the standard model Higgs sector,
but, more importantly, our experience in physics tells us that the physics 
responsible for 
electroweak symmetry breaking will, at the very least, allow a description of 
the mass scale in terms of a symmetry.

What will this new symmetry be?
There are many possibilities, but it is useful to group them according to 
the fate of the hypothetical Higgs boson. 
There are three logical possibilities
\begin{enumerate}
\item There is no Higgs boson.
\item The Higgs boson is composite (at a scale close to the weak scale).
\item The Higgs boson is elementary.
\end{enumerate}

The first option is realized in technicolor theories where
the weak scale arises by dimensional transmutation
from a gauge coupling, just like in QCD. 
The second option can also be realized by having a new strong gauge
force.
In this case the new strong force first produces a composite scalar
bound state, which then becomes the Higgs boson of electroweak symmetry 
breaking.
In both these examples, the symmetry description of the weak scale is in 
terms of the symmetry group of some new gauge force.

The third option is quite different.
The only known symmetry description for a fundamental Higgs boson involves 
supersymmetry. The lightness of the Higgs may be related to a chiral symmetry 
acting on its fermionic 
superpartner, or it may be due to the Higgs being a pseudo-Goldstone boson.
In either case, the weak scale is the scale at which supersymmetry is broken. 
To get a deeper understanding of the weak scale one must then address the 
question of how supersymmetry is broken.
Presumably, the reason for why the weak scale is much less than the Planck 
scale is the same as for the technicolor and composite Higgs options: 
it occurs as a dimensional transmutation due to the strong dynamics of a new 
interaction.
Whereas in the technicolor case one can simply appeal to the analogy with QCD,
in the supersymmetry case there is no analogy - nature has not provided us 
with other examples of broken spacetime symmetries
- hence there is no substitute for understanding the dynamics of the field 
theory.

\medskip

\noindent{\bf II.2 Matter v. Higgs}

\medskip

In the standard model it is obvious what distinguishes matter fields,
the quarks and leptons, from the Higgs field:
matter fields are fermions, while Higgs fields are bosons.
In supersymmetry this distinction disappears!
Once superpartners are added, there is no spacetime distinction
between quarks, $(q, \widetilde{q})$, leptons $(\ell, \widetilde{\ell})$
and Higgs $(\widetilde{h}, h)$ supermultiplets, both contain a fermion 
$(q, \ell$ or $\widetilde{h})$ and a boson $(\widetilde{q}, \widetilde{\ell}$ 
or $h$).
Indeed, the distinction between the lepton doublet and the Higgs doublet 
becomes a puzzle of fundamental importance.
Since these have the same gauge quantum members, what is the theoretical 
distinction between the Higgs and the lepton superfield?

Supersymmetry apparently allows us to do without a Higgs supermultiplet:
why not identify the Higgs boson with one of the sneutrino fields, 
$\widetilde{\nu}$?
If there are three generations of matter then this is not possible: a 
sneutrino vev $\vev{\widetilde{\nu}}$ leads to a Dirac mass of size $M_Z$ 
coupling the correspond $\nu$ state to the $\widetilde{Z}$.
Such a theory would only have two neutrinos of mass less than $M_Z$.
The sneutrino as Higgs idea is so attractive, that it is worth considering
the Higgs to be the sneutrino of a fourth generation.
In this case it is the fourth neutrino which marries the $\widetilde{Z}$ to 
acquire mass $M_Z$, which has the added advantage of explaining why only three
neutrinos are seen in the $Z$ width.
The problem with this scheme is that supersymmetry forbids a tree-level 
coupling of the sneutrino to the up type quarks:
the $t$ and $t'$ masses would have to occur via radiative corrections.
Given these large masses, this would necessarily involve new non-perturbative 
interactions. With just four generations of chiral superfields, and the known
gauge interactions, the only interactions which could break the chiral symmetry
on $u_R$ is the trilinear scalar interaction $\tilde{q} \tilde{u}
\tilde{\ell}^\dagger$. Such non-holomorphic supersymmetry breaking interactions
are not usually considered $-$ however, they do not introduce quadratic
divergences. This interaction is asymptotically free, so that it
could become non-perturbative at low energies. However, it is very unclear
whether it could give rise to sufficiently large masses for $t$ and $t'$
quarks.

Perhaps the above line of reasoning has not been developed further because the
unification of gauge couplings in supersymmetric theories suggests that
there are two light Higgs supermultiplets at the weak scale which are distinct
from the matter.
The conventional picture of weak scale supersymmetry 
has Higgs superfields, $h_1$ and $h_2$, which are distinct
from the lepton superfields, although the origin of the distinction indicates 
that there must be yet another symmetry. The nature of this symmetry is
discussed in III.7.

\vskip .2in

\noindent {\bf II.3 A heavy top quark effect}

\medskip

As mentioned in II.1, supersymmetry is the only known tool that allows a
fundamental Higgs boson at the electroweak scale to be understood in terms of
symmetries. This understanding has two aspects

$\bullet$ The size of $|m_h^2|$ is controlled by the scale of supersymmetry
breaking, which is presumably determined by some strong dynamics leading to a
dimensional transmutation. Candidate field theories for this exist, but we are
far from having a standard picture for the origin of supersymmetry breaking, 
and I will not discuss it further in these lectures.

$\bullet$ The sign of  $m_h^2$ is controlled by the dynamics which connects the
particles of the standard model to the supersymmetry breaking interactions, and
also by radiative corrections to  $m_h^2$. A
given model makes this dynamics explicit, and, if it is perturbative, the sign
of  $m_h^2$ is calculable.

In the most popular schemes for giving mass to the superpartners, the
supergravity and gauge messenger schemes mentioned in I.5, the messenger
dynamics is perturbative and leads to positive mass squareds for all scalars in
the theory. This makes the issue of how $SU(2) \times U(1)$ breaks, ie of why
$m_h^2$ is negative, particularly pressing. In particular, what distinguishes
the Higgs boson from the other scalars in the theory, the scalar quarks and
leptons, which must have positive mass squareds?

The answer to this puzzle is made plausible by its simplicity.
There are two important radiative corrections to any scalar mass, $m^2$

$\bullet$ gauge contributions, which increase $m^2$, and

$\bullet$ Yukawa contributions, which typically decrease $m^2$.

The only important Yukawa radiative corrections are induced by the large top 
Yukawa coupling $\lambda_t$ \footnote{The $b$ and $\tau$ Yukawa couplings could
also be large, in which case the conclusions of this section
are strengthened.}. Hence all $m^2$ are kept positive by the gauge radiative
corrections, with the possible exceptions of $m_h^2$ and $m_{\tilde{t}}^2$, 
since only $h$ and $\tilde{t}$ couple to $\lambda_t$. The $\lambda_t^2$
radiative correction is more powerful for $m_h^2$ than for $m_{\tilde{t}}^2$,
meaning that it is  $m_h^2$ which has the greater tendency to go negative. This
is due to the colored triplets have a larger multiplicity that weak doublets:
$SU(2)$ breaks rather than $SU(3)$ because it is a smaller group. Once  $m_h^2$
is negative the Yukawa corrections to $m_{\tilde{t}}^2$ actually change sign,
preventing $m_{\tilde{t}}^2$ from becoming negative.
In addition, $m_{\tilde{t}}^2$, has QCD radiative 
corrections which also make it more positive than $m_h^2$.

Electroweak symmetry breaking is therefore understood to be a large top quark
mass effect; a result which was obtained before the top quark was known to be
very heavy \cite{IR2,ACW}. Keeping other parameters of the theory fixed,
$\lambda_t$ is the order parameter for electroweak symmetry breaking in
supersymmetric models. For low values of $\lambda_t$, $SU(2) \times U(1)$ is
unbroken, whereas for high values of $\lambda_t$ it is broken. The critical
value for $\lambda_t$ does depend on other parameters of the theory, for
example the superpartner masses. However, now that we know that the top quark
is about 175 GeV, $\lambda_t$ is above the critical value for a very wide range
of parameters. I am tempted to say that electroweak symmetry breaking is hard
to avoid, but such a statement would require a detailed numerical study.

The size of $|m_h^2|$, and therefore $M_Z$, and the superpartner masses are
both determined by the scale of supersymmetry breaking. Does this allow a
prediction of the masses of the superpartners? Since there is more than one
supersymmetry breaking parameter, the answer is no. Nevertheless, the
understanding of the weak scale from symmetry principles requires that the
superpartners not be much heavier than $M_Z$. Denote the set of supersymmetry
breaking parameters by the scale $m_s$ and the dimensionless parameters $a$.
For example, $m_s$ could be defined to be the mass of the lightest chargino, 
and one of the $a$ parameters would be the ratio of the top squark mass to this
chargino mass. Since $M_Z$
has its origin in supersymmetry breaking, it is necessarily given by a formula
of the form $M_Z^2 = m_s^2 f(a)$. The scale of the superpartner masses, $m_s$,
can be made much larger than $M_Z$ only at the expense of a fine tuning amongst
the $a$ parameters to make $f(a)$ small. Hence

$\bullet$ We cannot predict the mass of the superpartners. (Certain superpartner
mass ratios are predicted in given messenger schemes, and in certain theories
with flavor symmetries, and are important tests of these theories.)

$\bullet$ The superpartner mass scale, $m_s$, can be made much larger than $M_Z$
only by a fine tune between dimensionless parameters which increases as
$m_s^2/M_Z^2$. 

The amount of fine tuning can be characterized by the sensitivity of $M_Z^2$
to small changes in the $a$ parameters:
$c_a = (a/M_Z^2) \delta M_Z^2/\delta a$ \cite{BG}.
A refined definition of the sensitivity parameter, $\gamma_a = c_a/ \bar{c}_a$, 
has been advocated, where $\bar{c}_a$ is an average of $c_a$ \cite{AC}.
Although there are no rigorous, mathematical upper bounds 
on the superpartner masses, it is possible to give upper bounds on the
superpartner masses if the amount of fine tuning, taken to be  
$\tilde{\gamma}$, the largest of the $\gamma_a$, 
is restricted to be less than a certain value. Such
naturalness bounds are shown for the Higgs scalar masses as well as the 
superpartner masses in the Figure. The upper extent of the line corresponds to 
$\tilde{\gamma} = 10$, the error bar symbol to  $\tilde{\gamma} = 5$, and the
squares give values of the masses for which the fine tuning is minimized. This
plot applies to the case of universal boundary conditions on the scalar masses
at very high energies. Relaxing this condition will allow some superpartner 
masses, for example the scalars of the first two generations, to increase
substantially. However, there will still be several superpartners, such as 
the lighter charginos ($\chi^+$), the lighter neutralinos ($\chi^0$), and 
the top squarks, which will prefer to be lighter than 300 GeV. The absence of
any superpartners beneath 1 TeV would mean that the understanding of the weak
scale described in this chapter has very serious problems. 
LEP II and the Fermilab Main Injector are well
positioned to discover supersymmetry, although the absence of superpartners at
these machines would not be conclusive.

\newpage

{\large \bf III. Flavor in Supersymmetric Theories.}

\noindent {\bf III.1 The fermion mass and flavor changing problems}

\medskip

In nature fermions exist in 45 different helicity states. What is the origin of
these states, and why do they assemble into three generations of quarks and
leptons with such diverse masses, mixings, gauge and global quantum numbers?
This is the flavor problem. Two important aspects of the flavor problem are:

(1). The fermion mass problem. What is the origin of the observed hierarchy of
quark and lepton masses and mixings?

Models of particle physics can be divided into two groups. {\it Descriptive
Models} are those which describe the observed quark and lepton masses and 
mixings
with 13 free parameters and make no attempt to understand the hierarchies. The
standard model is a descriptive model.
{\it Predictive Models} are those which either describe the 13 observed masses
and mixings with fewer than 13 parameters, or which provide some understanding
of the mass and mixing angle hierarchies.

(2). The flavor-changing problem. Why are processes which involve
flavor-changing neutral 
currents (FCNC) so rare? Three such highly suppressed quantities are 
$\Delta m_K, \epsilon_K$ and the rate for $\mu \to e\gamma$.

Coupling constants which distinguish between generations are called flavor 
parameters, 
and include the parameters which generate the observed quark and lepton 
masses and mixing.
In the standard model there are 13 flavor parameters, precisely
one for each of the 13 observed fermion masses and mixings, and
they all originate from the Yukawa coupling matrices.
In extensions of the standard model there may be more flavor parameters, 
so that they cannot all be experimentally determined from the quark and 
lepton masses and mixings.

A model is considered natural if it suppresses FCNC processes 
for {\it generic} values of the flavor parameters, ie for a wide range of the
parameters that is consistent with the observed fermion masses and mixing.
The standard model is natural in
this sense: all the Yukawa parameters are determined from the experimentally
measured fermion masses and mixings, and the GIM mechanism \cite{GIM} ensures 
the smallness of FCNC processes. For  models 
with more  flavor parameters we must address the question
of what values of the parameters are generic.

In this chapter, I assume that below some high scale $\Lambda$, physics is
described by a softly-broken, supersymmetric $SU(3) \times SU(2) \times U(1)$
gauge theory of minimal field content: three generations of quark and lepton
superfields $q_i, u_i, d_i, l_i$ and $e_i$ and two Higgs doublet superfields
$h_1$ and $h_2$ Assuming invariance under $R$ parity, the flavor parameters of
this theory can be written as 11 matrices in generation space.
Three of these are Yukawa coupling matrices of the superpotential
$$
W = q \mbox{\boldmath$\lambda$}_U u h_2 + q \mbox{\boldmath$\lambda$}_D 
d h_1 + \ell \mbox{\boldmath$\lambda$}_E e
h_1.\eqno(III.1)
$$
The supersymmetric interactions have identical flavor structure to the standard
model, and lead to a supersymmetric GIM mechanism suppressing FCNC effects. The
other 8 matrices contain soft supersymmetry breaking parameters
$$
\eqalignno{
V_{soft} &= \tilde{q} \mbox{\boldmath$\xi$}_U \tilde{u} h_2 + 
\tilde{q} \mbox{\boldmath$\xi$}_D\tilde{d} h,
+ \tilde{\ell} \mbox{\boldmath$\xi$}_E\tilde{e} h_1 + h.c.\cr
&+ \tilde{q} {\bf{m}}^2_q\tilde{q}^\dagger + \tilde{u}^\dagger
{\bf{m}}^2_u \tilde{u} +
\tilde{d}^\dagger {\bf{m}}^2_d\tilde{u} + \tilde{\ell}
{\bf{m}}^2_\ell\tilde{\ell}^\dagger +
\tilde{e}^\dagger{\bf{m}}^2_e\tilde{e}&(III.2)\cr}
$$
If these 8 matrices are given values which are ``generic'', that is the 
size of any 
entry in a matrix is comparable to the size of any other entry, then 
loop diagrams
involving superpartners lead to very large FCNC effects, even for 
superpartners as heavy as 1 TeV\cite{DG}.
For example the quantities $\epsilon_K$ and $\Gamma (\mu \to e\gamma)$ are 
about 10$^7$ larger than allowed by experiment.
This is the flavor-changing problem of supersymmetry.

Over the last few years an interesting new development has occured.
Progress has been made simultaneously on the fermion mass and
flavor changing problems of supersymmetry by introducing flavor symmetries
which constrain the forms of both the Yukawa couplings of (III.1) and the scalar
masses and interactions of (III.2). In the symmetry limit, many of the Yukawa
coupling entries vanish, and the form of the scalar masses are strongly
constrained. Small hierarchical breakings of the flavor symmetry introduce
small parameters that govern both the small masses and mixings of the fermions,
and the small violations of the superGIM mechanism which 
give small contributions to FCNC
processes. This linking of two problems is elegant and constraining; it is so
simple that it is hard to understand why it was not explored in the early
eighties. Perhaps we are taking supersymmetry more seriously these days.

In section III.5 I will discuss the literature on this subject, which began in
1990 and has grown into a minor industry recently. 
Each of the papers to date studies a particular
flavor symmetry $G_f$ and a particular breaking pattern. Many of the models
illustrate a special point or aim for a particular fermion mass 
prediction. In sections III.2 and III.3 below, my aim is to demonstrate the
generality and power of this approach. In fact from this viewpoint, I argue that
the flavor changing problem has arisen because of an unreasonable 
definition of ``generic.''
We know from the observed masses and mixings of quarks that 
$\lambda_{D_{12}}$ and $\lambda_{D_{21}}$ are very small.
A solution to the fermion mass problem would give us an understanding of why
this is so, but no matter what the understanding,
the flavor symmetries acting on the down and strange quarks are 
broken only very weakly.
Experiment has taught us that approximate flavor symmetries (AFS) are a 
crucial aspect of flavor physics.
It is therefore quite unreasonable to take $m^2_{q_{12}} \approx 
m^2_{q_{11}}$; the 
former breaks strange and down flavor symmetries and hence should be very 
suppressed compared to the latter, which does not.
(A crucial difference between scalar and fermion mass matrices is that the 
diagonal entries of fermion mass matrices break Abelian flavor symmetries, 
while diagonal entries of scalar mass matrices do not.)

In this chapter, I explore the consequences of linking the flavor-changing 
problem to the fermion mass problem.
I require that {\it  all flavor parameters of the theory are
subject to the same approximate flavor symmetries}.
I take this to be an improved meaning of the word ``generic'' in the statement
of the flavor-changing problem.
With this new viewpoint it could be that there is no flavor-changing problem 
in supersymmetry.
Perhaps if one writes down the most generic soft parameters at scale 
$\Lambda$, the FCNC processes are sufficiently suppressed.

Let $G_f$ be the approximate flavor symmetry group of the theory below scale 
$\Lambda$, 
and suppose that $G_f$ is explicitly broken by some set of parameters 
\{$\epsilon (R)$\}, which transform
as some representation $R$ of $G_f$, and take values which lead naturally to the
observed pattern of fermion masses and mixings.
We will discover that for some $G_f$ and $\{\epsilon (R)\}$ the flavor 
problem is solved, 
while for others it is not. Hence the flavor-changing problem of supersymmetry
is transformed into understanding the origin of those 
$G_f$ and $\{\epsilon (R)\}$ which yield natural theories.

Below scale $\Lambda$, models are typically (but not always) {\it descriptive};
they do not provide an understanding of the fermion masses.
However, knowing which $G_f,\{\epsilon (R)\}$ solve the flavor changing 
problem serves as a guide to building {\it predictive} models above $\Lambda$.
The theory above $\Lambda$ should possess an exact flavor symmetry $G_f$ that 
is 
broken spontaneously by fields $\{\phi\}$, which transform  as $R$ under $G_f$
and have vacuum expectation values $\vev{\phi} = \epsilon \Lambda$.

In section III.2 I introduce the ideas of Approximate Flavor Symmetries (AFS),
and in section III.3 I give a set of simple conditions which are sufficient for
an AFS to solve the flavor changing-problem.
In section III.4 I show that the flavor changing problem is solved when $G_f$
is taken to be the maximal flavor symmetry.
I delay a discussion of previous work on this subject until section III.5.
In III.6 I discuss the case $G_f = U(2)$, where the flavor changing constraints
dictate a special and interesting texture for the fermion mass matrices. 
In III.7 I show that $R$ parity finds a natural home as a subgroup of the
flavor symmetry. Sections III.5 and III.7 are taken from \cite{CHM2}.
This chapter is the most technical of these lectures; a brief statement of the
conclusions is given in section III.8.

\noindent {\bf III.2 Approximate Flavor Symmetries.}

\medskip

Using approximate flavor symmetries to describe the breaking of flavor is hardly new, 
but it is certainly powerful.
QCD with three flavors has an approximate  flavor symmetry
$G_f = SU(3)_L\times SU(3)_R$, explicitly broken by various parameter $\{\epsilon (R)\}$, 
which include the quark mass matrix $M(3,\bar{3})$ and electric charge matrices $Q_L(8, 1)$ and $Q_R(1,8)$.
Below $\Lambda_{QCD}$ the flavor symmetries are spontaneously broken to
the vector subgroup  and $G_f$ is realized non-linearly.
The interactions of the Goldstone bosons can be described by constructing an invariant chiral
Lagrangian $({\cal{L}})$ for $\Sigma (3,\bar{3}) = exp (2i\pi/f)$.
For our purposes the crucial point is that the flavor symmetry breaking beneath 
$\Lambda_{QCD}$ can be described by constructing the
chiral Lagrangian to be a perturbation series in the breaking parameters 
$\{\epsilon\} = \{M, Q_L, Q_R ...\}$.
Thus
 $
{\cal{L}} = {\cal{L}}_0 + {\cal{L}}_1 + {\cal{L}}_2 + ...$ where ${\cal{L}}_N$
 contains terms of order $\epsilon^N$.
For example
$$
{\cal{L}}_1  = a_1 \Lambda^3_{QCD} Tr(M\Sigma^\dagger) + .....
\eqno(III.3a)
$$

$$
{\cal{L}}_2 = a_2 \Lambda^2_{QCD} Tr(M \Sigma^\dagger M \Sigma^\dagger) + 
a_3\Lambda^4_{QCD} Tr(Q_L \Sigma Q_R\Sigma^\dagger)+ ...\eqno(III.3b)
$$
where all the unknown dynamics of QCD appear in the
set of dimensionless strong interaction parameters
$\{a\}$, which are $O(1)$.
This illustrates the basic tool which we use in this chapter.

The full flavor symmetry of the 45 fermions of the standard model is $U(45)$.
This is broken to the group $U(3)^5$ by the standard model gauge interactions.
Each $U(3)$ acts in the 3 dimensional generation space, and is labeled by $A$,
which runs over the 5 types of fermion representation $(q, u, d, \ell, e)$.

The $U(3)^5$ flavor symmetry of the standard model gauge interactions is 
broken explicitly by
the Yukawa couplings of the standard model, which have the transformation 
properties
$$
\eqalignno{
\mbox{\boldmath$\lambda$}_U &(\bar{3}, \bar{3}, 1, 1, 1)\cr
\mbox{\boldmath$\lambda$}_D &(\bar{3}, 1, \bar{3}, 1, 1)\cr
\mbox{\boldmath$\lambda$}_E &(1, 1, 1, \bar{3}, \bar{3}).&(III.4)\cr}
$$
In this section we speculate that these Yukawa parameters result from some new
physics above scale $\Lambda$, which possesses an AFS $G_f$, broken 
explicitly by a set of parameters $\{\epsilon (R)\}$.
The theory beneath $\Lambda$ can be written as a perturbation series in 
the $\epsilon$.
The standard model gauge Lagrangian appears at zeroth order, while the 
flavor violating fermion masses appear at higher order.

Such a picture is not new: the composite technicolor standard models were 
based on this picture \cite{CG}.
In this case the theory above $\Lambda$ was taken to be a preonic theory with 
strong dynamics which leaves a $U(3)^5$ flavor symmetry unbroken.
The strong dynamics produces composite quarks, leptons and Higgs bosons.
The preonic theory contains parameters $\{\epsilon (R)\}$ which explicitly 
break $U(3)^5$; in fact these parameters are assumed to be preon mass matrices
${\bf{M}}_{U,D,E}$ with the same transformation 
properties as $\mbox{\boldmath$\lambda$}_{U,D,E}$. At first order in 
perturbation theory 
$\mbox{\boldmath$\lambda$}_{U,D,E}$ are generated proportional to 
${\bf{M}}_{U,D,E}$.
At higher order various phenomenologically interesting 4 quark and 4 
lepton operators are generated.
For example, the operator $1/\Lambda^6 (q {\bf{M}}_U{\bf{M}}^\dagger_Uq)
(q{\bf{M}}_U{\bf{M}}^\dagger_Uq)$ leads to an additional contribution to 
$\epsilon_K$.

This picture is very close to that adopted here, except that

(a) The theory beneath $\Lambda$ is one with softly broken supersymmetry, and 
contains 8
flavor matrices in the soft supersymmetry breaking interactions in addition 
to the three supersymmetric Yukawa matrices.

(b) A large variety of AFS groups $G_f$ and explicit symmetry 
breaking parameters $\{\epsilon (R)\}$ are of interest.
In this III.4 we consider the obvious possibility that 
$G_f = G_{max} = U(3)^5$, and 
$\{\epsilon (R)\} = \mbox{\boldmath$\epsilon$}_U, \mbox{\boldmath$\epsilon$}_D,
\mbox{\boldmath$\epsilon$}_E$
transforming as $\mbox{\boldmath$\lambda$}_{U,D,E}$ are the only symmetry 
breaking parameters.

(c) The more fundamental theory above $\Lambda$
 need not involve strong, non-perturbative dynamics.
Each possible term in the low energy theory will be given an arbitrary 
dimensionless coefficient (labelled by $\{ a\}$),
which we think of as being $O(1)$  if the dynamics at $\Lambda$ is strong.
However, if the dynamics at $\Lambda$ is perturbative,
 then $\{a\}$ will be less than unity, and the flavor-changing effects will
be milder.

As a final example of the previous use of AFS, we consider the standard model 
extended to contain several Higgs doublets.
It was frequently argued that these theories had a flavor-changing problem.
Those doublets orthogonal to the one with a vev could have Yukawa matrices 
unconstrained by fermion masses.
With all such couplings of order unity, the tree-level exchange of such
Higgs bosons generates large FCNC for
fermion interactions, such as $(1/m^2_h) (q_1 d_2)^2$ for 
$\Delta m_K$ and $\epsilon_K$.
For theories with several Higgs doublets at the weak scale, this flavor 
problem was 
frequently solved by imposing a discrete symmetry which allowed only a 
single Higgs to couple to the $u_i$ 
and only a single Higgs to the $d_i$ quarks\cite{GW}.

From the viewpoint of AFS, however, such discrete symmetries are
unnecessary \cite{AHR,HW}.
Suppose the Higgs doublet which acquires a vev is labelled $h_1$.
The hierarchical pattern of quark masses implies that the Yukawa interactions 
of $h_1$ possess an AFS.
It is unreasonable that $h_{2,3...}$ should have interactions which are all 
$O(1)$ and are unconstrained by these AFS. If one set of interactions 
possesses an AFS it is only natural that the entire theory is constrained by 
the same AFS. One possibility is that the AFS of the quark sector 
$G_Q = U(1)^9$, a $U(1)$ factor for each of 
$q_i, u_i$ and $d_i$\cite{AHR,HW}, with each $U(1)$ having its own symmetry
breaking parameter: thus $\epsilon_{q_i}$ transforms under $U(1)_{q_i}$ 
but not under any other $U(1)$, etc.
In this case all Yukawa couplings of $h_a$ to up quarks would have the 
structure 
$(\lambda^a_U)_{ij} \approx \epsilon_{q_i} \epsilon_{u_j}$ and to down quarks 
$(\lambda^a_D)_{ij}\approx \epsilon_{q_i}\epsilon_{dj}$.
The nine parameters $\{ \epsilon_{q_i}, \epsilon_{u_i}, \epsilon_{d_i}\}$ 
can be estimated from the six quark masses and the three 
Euler angles of the Kobayashi-Maskawa matrix. 
The flavor-changing problem of these multi-Higgs models is solved by
 such a choice of AFS, if the masses of the additional
scalars are several hundred GeV.
This simple Abelian symmetry is insufficient to solve the supersymmetric flavor
changing problem. It provides for no approximate degeneracy between
$\tilde{d}$ and $\tilde{s}$ 
squarks, and allows Cabibbo sized mixing between them, which, as shown in
the next section, leads to a disastrously large contribution to $\Delta M_K$. 

\medskip

\medskip

\noindent {\bf III.3 The flavor-changing constraints}

\medskip

A brief, somewhat heuristic, view of the general conditions required to solve 
the supersymmetric flavor changing problem will be given in this section.
The results will allow us to understand whether AFSs are likely to be of use in
solving this problem. 
My aim is to provide a set of sufficient conditions which I find to be
both simple and useful; I do not attempt to determine the necessary conditions.

Consider the case when $\mbox{\boldmath$\xi$}_{U,D,E} =O$.
Unitary transformations are performed on the fermion fields to diagonalize 
$\mbox{\boldmath$\lambda$}_{U,D,E}$ and on the scalar fields to diagonalize 
${\bf{m}}^2_a, a = q,u,d,\ell, e$.
In this {\it mass basis} there will be unitary mixing matrices at the 
gaugino vertices, which for the neutral gauginos we write as
$\bf{W}^\alpha$ where $\alpha = u_L, u_R, d_L, d_R, e_L, e_R$. Flavor
and CP violating effects are induced by Feynman diagrams involving 
internal gauginos and scalar superpartners.
These are box diagrams for $\Delta m_K, \epsilon_K, \Delta m_B$ ... and 
penguin-type diagrams for $\mu \to e\gamma, d_e, b \to s\gamma$ ... .
The exchange of a scalar of generation $k$ between external fermions (of 
given $\alpha$) of generations $i$ and $j$ leads 
to a factor in the amplitude of 
$$
X^\alpha_{ij} = m^2_s \sum_k W^\alpha_{ki} W^{\alpha*}_{kj}
P^\alpha_k\eqno(III.5)
$$
where $P_k^\alpha$ is the propagator for the scalar of mass $m^{\alpha}_k$.
$X^\alpha$ is made dimensionless by inserting a factor $m^2_s$, where $m_s$ 
describes the scale of supersymmetry breaking.
Studies of flavor and CP violating processes allows bounds to be placed on 
the magnitudes and imaginary parts of $X^\alpha_{ij}$ of the form
$$
X^\alpha_{ij}\  \ltap \ X^\alpha_{oij} \left({m_s\over m_{so}}\right)^P
\eqno(III.6)
$$
where the bound is $X_0$ when $m_s$ is taken to be the reference value $m_{so}$.
The quantity $p$ is a positive integer, so that the bounds become 
weaker for higher $m_s$.
For box diagram contributions $p=1$, while for penguin-like diagrams $p=2$.
Useful results for these bounds are tabulated in \cite{HKT,GMS,ACH2}, 
as are references to earlier literature.
For our purposes we extract the following results:

If ${\bf{W}}^\alpha$ are ``KM-like'' that is if 
$$
|W_{ij}^\alpha| \ \ltap \ |V_{ij}| (i \neq j).\eqno(III.7)
$$
where ${\bf{V}}$ is the Kobayashi-Maskawa matrix, important limits only 
result for processes where the external fermions
are of the first two generations (ie., neither $i$ nor $j$ is 3).

The most important flavor changing limits arise when $(i,j) = (1,2)$. For
example, taking the relevant phases to be of order unity, $\epsilon_K$ implies
$$
|X^\alpha_{12}| = m^2_s |W^\alpha_{21} W^{\alpha *}_{22} 
(P^\alpha_2-P^\alpha_1) + W^\alpha_{31}W^{\alpha *}_{32}
 (P^\alpha_3 - P^\alpha_1)| \ \ltap \ 10^{-4}.\eqno(III.8)
$$
Here and below I take $m_s = 1$ TeV.
For ${\bf{W}}^\alpha$ KM-like,
$\abs{W^\alpha_{31} W^{\alpha *}_{32}} \ltap 
\abs{V_{td}}\abs{V_{ts}}\approx 4\times 10^{-4}$,
so there is no constraint from the last term of equation (8) even if there 
is large non-degeneracy between the scalars of the first
and third generation.
It is the first term which is typically the origin of the supersymmetric 
flavor-changing problem. This first term I call the ``1--2" problem; while the
second term I call the ``1,2--3" signature, since if the ${\bf W}^\alpha$ are
CKM-like this contribution is close to the experimental value.
One way to solve the problem is to make $W^\alpha_{21}$ small
$$
\abs{W^\alpha_{21}} \ltap \abs{V_{td}} \abs{V_{ts}}.\eqno(III.9a)
$$
Another is to make the scalars $\tilde{\alpha}_1$ and $\tilde{\alpha}_2$ 
degenerate:
$$
\abs{D^\alpha_{21}} \ltap {\abs{V_{td}}\abs{V_{ts}}\over \abs{V_{us}}}\eqno(III.9b)
$$
where $D^\alpha_{ij} = (m^{\alpha^2}_i - m^{\alpha^2}_j)/m^{\alpha^2}_i$,
and in the limit of near degeneracy $D^\alpha_{12} \approx m^2_s(P^\alpha_2-
P^\alpha_1)$.
In fact, the condition (8) and  (9a) or (9b) need only be applied for 
$\alpha = d_L, d_R, e_L$ and $e_R$.
The limits to flavor-changing 
processes in the up sector are much weaker, and are not problematic. 
Of course, the flavor problem 
can also be solved by having smaller suppressions of both $W^\alpha_{21} $ 
and $D^\alpha_{21}$. Nevertheless,  I find it useful to keep in mind that,
for $\mbox{\boldmath$\xi$}_{U,D,E} = 0$, the flavor problem is solved if

I. All ${\bf{W}}^\alpha$ are KM-like.

II. Either (III.9a) or (III.9b) holds in the $d$ and $e$ sectors.

Since the $X^\alpha_{12}$ quantities are small, it is often convenient to
work in the  {\it gaugino basis}.
In this basis superfield unitary transformations are performed to diagonalize 
$\mbox{\boldmath$\lambda$}_{U,D,E}$ so that the neutral gaugino vertices are 
flavor conserving.
The scalar mass matrices now have off-diagonal entries which,
assuming they are small,  can be treated in perturbation
theory as flavor-violating interactions.
In this basis, (III.8) and (III.9a) or (III.9b) are replaced by
$$
\abs{{m^{\alpha^2}_{12}\over m^2_s}} \ \ltap \ 4 \times 10^{-4}.\eqno(III.9c)
$$

Until now we have avoided discussing the flavor matrices
$\mbox{\boldmath$\xi$}_{U,D,E}$ of equation III.1.
Inserting the Higgs vev induces mass mixing between left and right scalars,
hence 6 $\times $ 6 rotations are required to reach the mass basis.
It is easier to use the gaugino basis and treat these masses in perturbation 
theory, writing them as:
$$
\mbox{\boldmath$\xi$}_{U,D,E} = {\bf{W'}}^{u_L,d_L,e_L} 
\overline{\mbox{\boldmath$\xi$}}_{U,D,E} 
{\bf{W'}}^{u_R,d_R,e_R}\eqno(III.10)
$$
where $\overline{\mbox{\boldmath$\xi$}}_{U,D,E}$ are diagonal matrices. 
Experiments place many limits on the elements $\overline{\xi}_{U,D,Eii}$.
For our purposes it is useful to know that all these limits are satisfied if

III. All ${\bf{W'}}^\alpha$ are KM-like

IV. $\overline{\xi}_{U,D,Eii}$ are of order
 $m_s \overline{\lambda}_{U,D,Eii}$.\\
The basic reason for this is that the only large contributions to flavor 
changing processes involving the first two generations then
come from terms of order
$
|W'^{\alpha_L}_{31} W'^{\alpha_R}_{32}| \lambda_{b,t}\ \mbox{which are}\ \ltap 
\abs{V_{td}V_{ts}}.
$

Now that we have argued that the four statements (I) - (IV)
 are sufficient to solve the supersymmetric flavor problem, we can ask
whether it is reasonable to expect that AFS will be of use.
It should be apparent that the general expectation is that any AFS which
leads to the hierarchy of fermion masses, as parameterized by 
$\overline{\lambda}_{U,D,E_i}$,
and to the KM pattern of flavor violation, described by $V_{ij}$, 
will automatically lead to I, III 
and IV being satisfied.
The only remaining question is whether AFS can satisfy (II),
 ie., whether they can produce either (III.9a)
or (III.9b) (or (III.9c) in the insertion approximation).
The Abelian $G_f$ discussed earlier ($U(1)^9$ in the quark sector) is clearly
insufficient since it gives $D_{21}^\alpha \approx 1$ and $W_{21}^\alpha
\approx V_{us}$. In the next section I show that the maximal AFS is easily
sufficient.

\noindent{\bf III.4 The Maximal Approximate Flavor Symmetry.}

\medskip

We assume that below some high scale, $\Lambda$, physics is described by a 
softly broken, supersymmetric
$SU(3) \times SU(2) \times U(1)$ gauge theory with
minimal field content.
The flavor interactions are those of the superpotential and soft 
supersymmetry breaking interactions shown in equations (III.1) and (III.2).
We assume that the dynamics above $\Lambda$, which may be strong,
possesses an approximate flavor symmetry $G_f$.
Below $\Lambda$ the breaking of this AFS is characterized by a set of 
parameters $\{\epsilon (R)\}$ transforming as $R$ under $G_f$.
In this section we take $G_f$ to be $G_{max} = U(3)^5$, the maximal AFS which 
commutes with the standard model gauge group. 
Although strong dynamics could preserve a larger AFS, the breaking parameters
$\{\epsilon (R)\}$ cannot violate $SU(3) \times SU(2) \times U(1)$,
so that $G_{max}$ is the largest group under which the set $\{\epsilon\}$ 
form complete
 representations.
Each factor of $G_{max}$ is labelled as $U(3)_a$ where $a=q,u,d, l$ or $e$.
We assume that the $\{\epsilon\}$ fill
out three irreducible representations:
$\mbox{\boldmath$\epsilon$}_U \sim (3_q, \overline{3}_u), 
\mbox{\boldmath$\epsilon$}_D \sim
 (3_q, \overline{3}_d)$
and $\mbox{\boldmath$\epsilon$}_L \sim (3_l, \overline{3}_e)$.
In the case of QCD with approximate $SU(3)_L\times SU(3)_R$
broken explicitly by the quark mass matrix ${\bf{M}}$, there is no loss of 
generality in choosing
a basis for the quark fields in which ${\bf{M}}$ is real and diagonal.
Similarly, we may choose a basis for the lepton fields in which
$\mbox{\boldmath$\epsilon$}_E$ is real and diagonal
$\overline{\mbox{\boldmath$\epsilon$}}_E$.
We may choose the quark basis so that 
 $\mbox{\boldmath$\epsilon$}_U=\overline{\mbox{\boldmath$\epsilon$}}_U$
 is diagonal and $\mbox{\boldmath$\epsilon$}_D = 
{\bf{V}}^*\overline{\mbox{\boldmath$\epsilon$}}_D$, where 
$\overline{\mbox{\boldmath$\epsilon$}}_D$
is diagonal and ${\bf{V}}$ is a unitary matrix.
All flavor changing effects of this theory are described by a single matrix, 
which to high accuracy is the KM matrix. Criteria I and
III of the previous section are satisfied.
This theory has no violation of the lepton numbers.

To zeroth order in $\{\epsilon\}$, the only interactions of the quarks and 
leptons are the gauge interactions and the zeroth order
supersymmetry breaking potential
$$
V_0 = q m^2_q {\bf{1}} q^\dagger + u^\dagger m^2_u {\bf{1}} u + d^\dagger 
m^2_d {\bf{1}} d
+ \ell m^2_\ell {\bf{1}} \ell^\dagger + e^\dagger m^2_e {\bf{1}} e.\eqno(III.11)
$$
We see that the non-Abelian nature of $G_f$ enforces squark and slepton
degeneracy at zeroth order in $\epsilon$.
However, (III.11) differs from the universal boundary condition of supergravity 
because the five
parameters $m^2_a$ are all independent and are not constrained to be equal.
Similarly they can differ from the Higgs mass parameters. 
Equation (III.9b), and therefore criterion II, is satisfied at zeroth order,
but corrections appear at higher order.

At first order in $\epsilon$, superpotential interactions are generated:
$$
W_1 = a_1\; q \mbox{\boldmath$\epsilon$}_U uh_2 + a_2\; q
\mbox{\boldmath$\epsilon$}_D dh_1+ a_3\;
 \ell \mbox{\boldmath$\epsilon$}_L
 eh_1\eqno(III.12) 
$$
where $a_{1,2,3}$ are ``strong interaction" parameters of order unity.
The $U(3)$ transformations are shown explicitly in Appendix A at the end of this
chapter.

The assumed transformation properties of the $\{\epsilon\}$ are sufficient 
to guarantee that $W$ preserves $R$ parity invariance to
all orders in $\epsilon$.
There is no need to impose $R$ parity as a separate exact symmetry.
The Yukawa couplings can be written as expansions in $\epsilon$,
for example $\lambda_U = a_1 \epsilon_U + a_4 \epsilon_U\epsilon_U^\dagger
\epsilon_U + a_5 \epsilon_D\epsilon_D^\dagger \epsilon_U + ...$ .
If we work only to second order, we can simply take $\lambda_U = a_1 
\epsilon_U$, etc.
Even if we work to higher order, we can rearrange the perturbation series as 
an expansion in 
$\mbox{\boldmath$\lambda$}_{U,D,E}$ rather than 
$\mbox{\boldmath$\epsilon$}_{U,D,E}$.
Either way, to second order in the expansion:
$$
W_1 = q \mbox{\boldmath$\lambda$}_U u h_2 + 
q \mbox{\boldmath$\lambda$}_D dh_1 + \ell \mbox{\boldmath$\lambda$}_E eh_1
\eqno(III.13a)
$$
$$
W_2 = {a_1\over \Lambda^2} (q \mbox{\boldmath$\lambda$}_U u)
(q \mbox{\boldmath$\lambda$}_D d) + ...
\eqno(III.13b)
$$
$$
V_1 = m_s(a_U q \mbox{\boldmath$\lambda$}_U uh_2 + 
a_D q \mbox{\boldmath$\lambda$}_D dh_1 + a_E \ell 
\mbox{\boldmath$\lambda$}_E eh_1)\eqno(III.13c)
$$
$$
\eqalignno{
V_2 &= m^2_s \bigg( q(a_2 \mbox{\boldmath$\lambda$}_U
\mbox{\boldmath$\lambda$}_U^\dagger + a_3 
\mbox{\boldmath$\lambda$}_D
\mbox{\boldmath$\lambda$}_D^\dagger)q^\dagger
+a_4 d^\dagger \mbox{\boldmath$\lambda$}^\dagger_D
\mbox{\boldmath$\lambda$}_D d + a_5 u^\dagger
\mbox{\boldmath$\lambda$}^\dagger_U 
\mbox{\boldmath$\lambda$}_U u  + 
a_6 \ell \mbox{\boldmath$\lambda$}_E
\mbox{\boldmath$\lambda$}_E^\dagger\ell^\dagger \cr & + 
a_7 e^\dagger \mbox{\boldmath$\lambda$}^\dagger_E
\mbox{\boldmath$\lambda$}_Ee)
+ {m^2_s\over\Lambda^2} a_8 (q \mbox{\boldmath$\lambda$}_U
\mbox{\boldmath$\lambda$}_U^\dagger q^\dagger)
(u^\dagger u)+
{m^2_s\over \Lambda^2} a_9 (q \mbox{\boldmath$\lambda$}_U u)
(q \mbox{\boldmath$\lambda$}_D d).&(III.13d)\cr}
$$
Given the non-renormalization theorems, one might question whether the 
interactions in $W$ really are generated. 
In general the answer is yes: they are generated by integrating out heavy 
particles at tree level
and by radiative corrections to $D$ terms followed by field rescalings.
However, in specific simple models, one discovers that the structure of the
supersymmetric theory is such that not all interactions allowed by the
symmetries of the low energy theory are generated. Hence if the symmetry
structure of the low energy theory is insufficient to solve the flavor changing
problem, it may still be that a theory above $\Lambda$ with this symmetry can 
be constructed which does not generate the troublesome interactions.

In QCD the strong interaction parameters are real -- the strong dynamics of QCD 
preserves CP.
Also, the strong dynamics is well separated from the origin of the explicit
breaking parameters 
$\mbox{\boldmath$\epsilon$} = {\bf{M}}, {\bf{Q}}$.
The ``strong'' dynamics of the supersymmetric theory above $\Lambda$ may 
conserve CP so that
$a_1 ... a_9$ are real.
This would explain the smallness of the neutron electric dipole moment which has
contributions from $Im (a_u)$ and $Im(a_d)$ \cite{EFN}.
However it may be that the dynamics above $\Lambda$ which generates these 
coefficients is not very separate from that which
generates the $\{\epsilon\}$. 
Since the KM phase comes from $\{\epsilon\}$, in this case there would also 
be phases in $\{a\}$.

Does the boundary condition of (III.11) and (III.13) at scale $\Lambda$ solve 
the flavor-changing problem? In the lepton sector the answer is obviously yes: 
$\mbox{\boldmath$\lambda$}_E$ can be made real and diagonal so there is 
no lepton  flavor violation.

In the quark sector the only mixing matrix is the KM matrix, so that criteria 
I and III are satisfied.
In fact, the only unitary transformations needed to reach the mass basis are 
a rotation of ${\bf{V}}$ on $d_L$ quarks, and a rotation of $q$ squarks.
This latter rotation is awkward; it is more convenient to make the ${\bf{V}}$ 
rotation on $d_L$
to be a superfield rotation, and to treat the remaining scalar mass flavor 
violation as a perturbation:
$$
{\delta{\bf{m}}^{d_L^2}_{21}\over m^2_s} = a_2 
({\bf{V}}^T \overline{\mbox{\boldmath$\lambda$}}^2_U {\bf{V}}^*)_{21} \approx
a_2 \abs{V_{ts}V_{td}}^* \lambda^2_t\approx 4 \times 10^{-4}.\eqno(III.14)
$$
We can see that the  condition (III.9c), and  therefore criterion II, 
is satisfied. 
Finally, the trilinear scalar interactions of $V_1$ 
in (III.13c) clearly satisfy the criterion IV.
The matrices ${\bf{W'}}^\alpha = {\bf{I}} + O(\mbox{\boldmath$\epsilon$}^2)$ 
so that III is also satisfied.

The flavor structure of this theory with $G_f = G_{max} = U(3)^5$ is
very similar to that which results from the universal boundary conditions
of supergravity discussed below.
In that theory the terms $a_2 ... a_9$ are assumed to be absent at the 
boundary, but are generated via renormalization groups scalings from 
$\Lambda = M_{Pl}$ to $m_s$, and end up being of order unity.
What features of this flavor sector are crucial to solving the flavor changing 
problem?

i) At zeroth order in $\epsilon$ the scalars of each $A$ 
are degenerate and the soft operators have no flavor violation.

ii) At linear order in $\epsilon$, the superfield rotations which diagonalize 
the quark masses also diagonalize the soft scalar trilinear couplings.
Hence at this order the soft operators contain no flavor changing neutral 
currents.

iii) The corrections to ${\bf{m}}^2_a$, induced at second order in $\epsilon$, 
induce flavor changing effects proportional to
$\mbox{\boldmath$\lambda$}_U
\mbox{\boldmath$\lambda$}_U^\dagger$ and $\mbox{\boldmath$\lambda$}_D
\mbox{\boldmath$\lambda$}_D^\dagger$.
If we restrict $\mbox{\boldmath$\lambda$}_U$ and 
$\mbox{\boldmath$\lambda$}_D$ to their light $2\times 2$ 
subspaces then all contributions are less than $10^{-4}$.
Hence we need only consider contributions involving the heavy generation.
For external light quarks this gives small contributions because $V_{ts}$
and $V_{td}$ are small.

We finish this section by briefly comparing the AFS method to several 
well-known solutions of the supersymmetric flavor-changing problem.
The low energy structure of these theories can be understood as examples of
the AFS technique.

The most popular treatment of the supersymmetric flavor-changing problem
is to assume that at some high scale, usually taken to be the 
reduced Planck mass, the flavor matrices possess a ``universal'' form
\cite{BFS,HLW}:

$$
{\bf{m}}^2_a = m^2_0 {\bf{I}}\eqno(III.15a)
$$
$$
\mbox{\boldmath$\xi$}_{U,D,E} = A \ \mbox{\boldmath$\lambda$}_{U,D,E}
\eqno(III.15b)
$$
which generalizes the idea of squark degeneracy
\cite{DG}.
This form is the most general which results from hidden sector
supergravity theories, provided the K\"ahler
potential is $U(N)$ invariant, where $N$ is the total number of chiral 
superfields \cite{HLW}.
However, imposing this $U(N)$ invariance as an exact symmetry on one piece 
of the Lagrangian is ad hoc because it is broken explicitly by the gauge
and superpotential interactions.

We advocate replacing this $U(N)$ idea with an approximate 
flavor symmetry $G_f$ acting on the entire theory, 
broken explicitly by a set of parameters $\{\epsilon (R)\}$, allowing the 
Lagrangian
to be written as a power series in $\epsilon$: ${\cal{L}}_0+{\cal{L}}_1+ ...$.
At each order the most general set of interactions is
written which is consistent with the assumed transformation properties of 
$\{ \epsilon (R)\}$. Taking $G=U(3)^5$ 
we have found that a modified universal boundary condition emerges.
At zeroth order in $\epsilon$ we found (III.15a) to replaced by 
$$
{\bf{m}}^2_a = m^2_a {\bf{I}}\eqno(III.16a)
$$
and at first order in $\epsilon$, (III.15b) is replaced by 
$$
\mbox{\boldmath$\xi$}_{U,D,E} = A_{U,D,E} \mbox{\boldmath$\lambda$}_{U,D,E}.
\eqno(III.16b)
$$
These boundary conditions are corrected at higher orders by
factors of $(1+O(\epsilon^2))$, but are sufficient to solve the supersymmetric
flavor-changing problem.
While (III.15) was invented as the most economical solution to the flavor 
changing problem, 
the symmetry structure of the theory demonstrates that it is ad hoc,
and from the phenomenological viewpoint it is overkill.
The flavor structure of the low energy theory provides a motivation 
for (III.16), together with the $1 + O(\epsilon^2)$ correction factors.
Phenomenological results, which follow from assuming the boundary condition (15)
but do not result from (III.16), should be considered suspect. 
For example, the flavor changing problem provides no motivation 
for the belief that the squarks of the lightest generation
$(\tilde{q}_L, \tilde{d}_R, \ \hbox{and}\ \tilde{u}_R)$ are degenerate
(up to electroweak renormalizations and breaking).
Similarly, the flavor changing problem provides no motivation for a boundary 
condition where 
$m^2_{h_1}$ and $m^2_{h_2}$ are  both set equal to squark and slepton masses.

Perhaps the most straightforward idea to solve the flavor changing problem
is to assume that supersymmetry breaking is transferred to the observable 
sector by the known gauge interactions \cite{ACW}.
Suppose this happens at scale $\Lambda$, and that below $\Lambda$ the observable
sector is the minimal field content supersymmetric $SU(3)\times SU(2)\times 
U(1)$ theory. At scale $\Lambda$ the dominant soft supersymmetry breaking
operators are the three gaugino mass terms, 
which are generated by  gauge mediation at the 1 loop level.
At higher loop level, at scale $\Lambda$ the eight flavor matrices
${\bf{m}}^2_a$ and $\mbox{\boldmath$\xi$}_{U,D,E}$ are generated.
However, since the only violation of the $U(3)^5$ flavor symmetry is provided
by
$\mbox{\boldmath$\lambda$}_{U,D,E}$, {\it the most general theory of this 
sort is described at scale 
$\Lambda$ by equations (III.11) and (III.13) and hence possesses the boundary 
condition (III.16)}.
The parameters $\{a\}$ are now each given by a power series in the standard 
model gauge couplings,
$\alpha_i$, with coefficients which depend on the representation structure 
of the supersymmetry breaking sector.
The gaugino masses $M_i$ are very large, and at 
low energy the parameters $m^2_A$ of (III.11) 
receive contributions $\propto \sum_i C_{iA} \alpha_i M^2_i \ln \Lambda/m_s$, 
where $C_{iA}$ involve quantum numbers.
This may dominate $m^2_a$ boosting the importance of $V_0$, and thereby
decreasing the flavor violating effects induced by $V_{1,2}$.

The AFS technique is sufficiently general that it can be used no matter how
supersymmetry is broken and transmitted to the observable sector. This almost
guarantees that it will be a useful tool in studying the flavor questions of
supersymmetry. It may be that nature chooses a more complicated $G_f$ and
{$\epsilon$} than the above example. At scale $\Lambda$ the observable sector 
may involve additional fields and there may be additional flavor breaking 
matrices. Simple group theory can be used to determine the additional terms
which these induce in $V_1$ and $V_2$, allowing an easy estimation of potential
flavor-changing difficulties.

In the previous section we argued that approximate flavor symmetries which 
lead to the observed 
hierarchy of quark and lepton masses and mixings are very likely to give 
supersymmetric theories where all mixing matrices are KM like, and the 
eigenvalues of 
$\mbox{\boldmath$\xi$}_{U,D,E}$ possess a hierarchy similar to the 
eigenvalues of 
$\mbox{\boldmath$\lambda$}_{U,D,E}$.
Hence the criteria I, III, and IV are easily satisfied, and the real flavor 
problem is that either (III.9a) or (III.9b) must be imposed.
This means that either the mixing in the first two generations, 
$W^\alpha_{21}$, is much 
smaller than expected from the Cabibbo  angle, or the squarks of the first 
two generations must be highly 
degenerate. This degeneracy can be understood as the consequence of a 
non-Abelian symmetry, 
continuous or discrete, which acts on the first two generations.
The low energy limit of any such theories can be analysed using AFS. An 
alternative possibility is to seek Abelian symmetries, allowing squark 
non-degeneracies, which lead to the suppression of $W^\alpha_{21}A$.

It is well-known that the experimental constraints on FCNC imply that 
$W^\alpha_{21}$ need be suppressed only in the $d$ and $e$ sectors 
$(\alpha = d_L, d_R, e_L, e_R)$:
$W^{u_L}_{21} \approx W^{u_R}_{21}\approx V_{us}$ leads to interesting 
$D^0\overline{D}^0$ mixing, but is not a problem.
This opens the possibility that symmetries can 
be arranged so that Cabibbo mixing originates in the $u$ sector, 
while mixing of the generations is highly suppressed in the $d$ and $e$ sectors.
This idea has been used to construct models with Abelian flavor symmetries 
and non-degenerate squarks \cite{NS}.

\vskip 9pt

{\bf III.5 A brief introduction to the literature.}

\vskip 9pt

In supersymmetric models of particle physics there are two aspects to the 
flavor problem. The first is the problem of quark and lepton mass and mixing 
hierarchies: why are 
there a set of small dimensionless Yukawa couplings in the theory?
The second aspect of the problem is why the superpartner gauge interactions 
do not violate flavor at too large a rate. This requires that the squark and 
slepton mass matrices not be arbitrary, rather, even though all eigenvalues 
are large, these matrices must also possess a set of 
small parameters which suppresses flavor-changing effects. 
What is the origin of this second set of small dimensionless parameters?

An extremely attractive hypothesis is to assume that the two sets of small
parameters, those in the fermion mass matrices and those in the scalar mass 
matrices, have a common origin: they are the small 
symmetry breaking parameters of an approximate flavor symmetry group $G_f$.
This provides a link between the fermion mass and flavor-changing problems; 
both are addressed by the same symmetry.
Such an approach was first advocated using a flavor group $U(3)^5$, broken 
only by the three Yukawa matrices $\lambda_{U,D,E}$ in the up, down and lepton
sectors \cite{HR}, as discussed in the previous section. 
This not only solved the flavor-changing problem, but 
suggested a boundary condition on the soft operators which has a more secure 
theoretical foundation than that of universality.
However, this framework did not provide a model for the origin
of the Yukawa matrices themselves, and left open the possibility that $G_f$
was more economical than the maximal flavor group allowed by the standard model
gauge interactions.

The first explicit models in which spontaneously broken flavor groups were 
used to constrain both fermion and scalar mass matrices were based on 
$G_f = SU(2)$ \cite{DKL} 
and $G_f = U(1)^3$ \cite{NS}.
In the first case the approximate degeneracy of scalars of the first two 
generations  was guaranteed by $SU(2)$. In retrospect it seems astonishing 
that the 
flavor-changing problem of supersymmetry was not solved by such a flavor
group earlier. The well known supersymmetric contributions to the $K_L - K_S$
mass difference can be rendered harmless by making the $\tilde{d}$ and 
$\tilde{s}$ squarks degenerate. Why not guarantee this degeneracy by placing
these squarks in a doublet of a non-Abelian flavor group 
$(\tilde{d}, \tilde{s})$? Perhaps one reason is that $SU(2)$ allows large
degenerate masses for $d$ and $s$ quarks.
In the case of Abelian $G_f$, the squarks are far from degenerate, however it 
was discovered that the flavor-changing problem could be solved by arranging
for the Kobayashi-Maskawa mixing matrix
to have an origin in the up sector rather than the down sector.

A variety of supersymmetric theories of flavor have followed, including ones 
based on $G_f =0(2)$ \cite{PoS}, $G_f=U(1)^3$ \cite{LNS}, $G_f=\Delta (75)$
\cite{KS}, $G_f=(S_3)^3$ \cite{HM,CHM,CHM2} and $G_f = U(2)$ \cite{PT,BDH}.
Progress has also been made on relating the small parameters of fermion and
scalar mass matrices using a gauged $U(1)$ flavor symmetry in a $N=1$
supergravity theory, taken as the low energy limit of superstring models
\cite{DPS}.
Development of these and other theories of flavor is of great
interest because they offer the hope that an understanding of the quark and 
lepton masses, and the masses of their scalar superpartners, may be
obtained at scales well beneath the Planck scale, 
using simple arguments about fundamental symmetries and how they are 
broken.
These theories, to varying degrees, provide an understanding of 
the patterns of the 
mass matrices, and may, in certain cases, also lead to very definite mass 
predictions. Furthermore, flavor symmetries may be of use to understand a 
variety of other important aspects of the theory.

The general class of theories which address both aspects of the supersymmetric 
flavor problem have
two crucial ingredients: the flavor group $G_f$ and the flavon fields, $\phi$,
which have a hierarchical set of vacuum expectation values allowing a sequential
breaking of $G_f$. These theories can be specified in two very different forms.
In the first form, the only fields in the theory beyond
$\phi$ are the light matter and Higgs fields.
An effective theory is constructed in which all gauge and $G_f$ invariant 
interactions are 
written down, including non-renormalizable operators scaled by some mass scale 
of flavor physics, $M_f$. An example of such a theory, with $G_f = U(3)^5$, was
discussed in section III.4.
The power of this approach is that considerable progress is 
apparently possible without 
having to make detailed assumptions about the physics at scale $M_f$
which generates the non-renormalizable operators.
Much, if not all, of the flavor structure of fermion and scalar masses comes 
from such non-renormalizable
interactions, and it is interesting to study how their form depends
only on $G_f$, $G_f$ breaking and the light field content.

A second, more ambitious, approach is to write a complete, renormalizable 
theory of flavor at the scale $M_f$.
Such a theory possesses a set of heavy fields which, when integrated out of 
the theory, lead to the effective theory discussed above \cite{FN}.
However, it is reasonable to question whether the effort required to construct 
such full theories is warranted.
Clearly these complete theories involve further assumptions beyond those of the
effective theories, namely the $G_f$ properties of the fields of mass
$M_f$, and it would seem that the low energy physics of flavor is independent 
of this, depending only on the properties of the effective theory.
In non-supersymmetric theories such a criticism may have some validity, but 
in supersymmetric theories it does not.
This is because in supersymmetric theories, on integrating out the states of 
mass $M_f$, the low energy theory 
may not be the most general effective theory based on flavor 
group $G_f$. Several operators which are $G_f$ invariant, and could be present 
in the effective theory, are typically not generated when the heavy states of 
mass $M_f$ are integrated out. Which operators are missing depends on what 
the complete theory at $G_f$ looks like.
This phenomena is well known, and is illustrated, for example, in 
references \cite{ADHRS,KS,BDH},
and it casts doubt on the effective theory approach to building supersymmetric 
theories of flavor. Finally, one might hope that a complete renormalizable 
theory of flavor at scale $M_f$ might possess a simplicity which is partly
hidden at the level of the effective theory.

\vskip 9pt

{\bf III.6 The Minimal U(2) Theory of Flavor.}

The largest flavor group which acts identically on each component of a
generation, and is therefore consistent with grand unification, is $U(3)$, with
the three generations forming a triplet. This is clearly strongly broken to
$U(2)$ by whatever generates the Yukawa coupling for the top quark. Hence the
largest such flavor group which can be used to understand the small parameters
of the fermion and scalar mass matrices is $U(2)$. In this section I briefly
mention aspects of the $U(2)$ theory constructed in reference \cite{BDH}.

While the third generation is a trivial $U(2)$ singlet, $\psi_3$, the two light
generations are doublets, $\psi_a$:
$$
q_a = \pmatrix{q_1\cr q_2} \; \; u_a = \pmatrix{u_1\cr u_2} \; \; d_a =
\pmatrix{d_1\cr d_2} \;
\; \ell_a =\pmatrix{\ell_1\cr \ell_2} \; \; e_a = \pmatrix{e_1\cr e_2}.
\eqno(III.17)
$$
In the symmetry limit only the fermions of the third generation have mass,
while the scalars of the first two generations are degenerate: clearly a
promising zeroth order structure.

The dominant breaking of $U(2)$ is assumed to occur via the vev of a doublet:
$\vev{\phi^a}$. If we study the most general theory beneath some flavor scale
$M_f$, then the non renormalizable operators for fermion masses are:
$$
{1 \over M_f} \; [\psi_3\phi^a\psi_a h]_F \eqno(III.18)
$$
which generates $V_{cb}$, and
$$
{1 \over M_f^2} \; [\psi_a\phi^a\phi^b\psi_b \; h]_F \eqno(III.19)
$$
which generates a 22 entry in the Yukawa matrices.
An immediate difficulty is that $U(2)$ also allows the supersymmetry breaking
scalar mass
$$
{1 \over M_f^2} \;[\psi^{\dagger a} \phi^\dagger_a \phi^b \psi_b \;
z^\dagger z]_D,\eqno(III.20)
$$
where $z$ is a supersymmetry breaking spurion, taken dimensionless,
$z=m \theta^2$, which leads to a splitting of the degeneracy of the scalar
masses of the first two generations:
$$
{m^2_{\tilde{e}} - m^2_{\tilde{\mu}}\over m_{\tilde{e}}^2 + m^2_{\tilde{\mu}}}
\approx O\pmatrix{m_\mu\over m_\tau}\eqno(III.21)
$$
in the lepton sector and
$$
{m^2_{\tilde{d}} - m^2_{\tilde{s}}\over m^2_{\tilde{d}} + m^2_{\tilde{s}}}
\approx O \pmatrix{m_s\over m_b}.\eqno(III.22)
$$
in the down quark sector. These lead to violations of the flavor changing 
constraints of section III.3 \cite{PT}. However, if 
these operators are generated by
Froggatt-Nielsen type theories \cite{FN}, one discovers that  III.21 and III.22
are not generated if the exchanged heavy vector generations transform as $U(2)$
doublets.

If the final breaking of $U(2)$ occurs via a two indexed antisymmetric tensor,
$\vev{A_{ab}}$ then the final operator contributing to fermion masses is
$$
{1 \over M_f} \; [\psi_a A^{ab}\psi_b h]_F \eqno(III.23)
$$
It is remarkable that theories of flavor can be based on the two interactions
of III.18 and III.23, in addition to the third generation coupling $[\psi_3
\psi_3 h]_F$. The Yukawa matrices take the form
$$
{\mbox{\boldmath$\lambda$}} = \pmatrix{ 0&\epsilon'&0\cr
-\epsilon'&0&\epsilon\cr
0&\epsilon &1}\eqno(III.24)
$$
where
$\epsilon = \vev{\phi^2}/M_f$ and $ \epsilon' = \vev{A^{12}}/M_f$,
and the scalar mass matrices are
$$
{\bf m}^2 = \pmatrix{ m^2_1 + \epsilon'^2 m^2 & 0 &\epsilon\epsilon' m^2\cr
0& m^2_1 + \epsilon'^2m^2 & 0\cr
\epsilon\epsilon' m^2 & 0 & m^2_3 + \epsilon^2 m^2}. \eqno(III.25)
$$  
The splitting between the masses of the scalars of the lightest two generations
is
$$
{m^2_{\tilde{e}} - m^2_{\tilde{\mu}}\over m_{\tilde{e}}^2 + m^2_{\tilde{\mu}}}
\approx O\pmatrix{m_e m_\mu^2 \over m_\tau^3}\eqno(III.26)
$$
in the lepton sector, with similar equations in the quark sector. The ``1-2"
aspect of the supersymmetric flavor changing problem is completely solved.
However, because $\lambda_{22}$ vanishes, the mixings to the third generation
are larger than those of the CKM matrix, so that the conditions of section 
III.3 are not immediately satisfied. 
The splittings between the third generation scalar mass and the
lightest two generations should not be of order unity, or the contribution to
$\epsilon_K$ from the ``12-3" effects in this model will be too large. This
splitting cannot be computed within a $U(2)$ theory, but will be an important
constraint on $U(3)$ theories.

This $U(2)$ theory of flavor has a significant economy of parameters. Two of
the standard model flavor parameters are predicted:
$$
\abs{ {V_{td}\over V_{ts}}} = s_1 = \sqrt{ {m_d\over m_s}} = 0.230 \pm 0.
008\eqno(III.27a)
$$
$$
\abs{ {V_{ub}\over V_{cb}}} = s_2 = \sqrt{ {m_u\over m_c}} = 0.063 \pm 0.
009\eqno(III.27b)
$$ 
As measurements of these quantities improve, it will be interesting to see
whether they remain within the uncertainties of the above predictions.
There are 6 unitary $3 \times 3$ flavor mixing matrices at neutralino vertices;
in the $U(2)$ theory they are real and given by 6 angles $s_{Iij}$ and 
$s_{Iij}^c$ where $I=U,D,E$ labels the up, down and lepton sectors, and $ij =
12,23,31$ labels the generations being mixed. These angles are predicted in
terms of just three free parameters $r_I$
$$
s_{I12} = -s_{I12}^c = \left( \sqrt{ {m_1\over m_2}}\right)_I\eqno(III.28a)
$$
$$
s_{I23} = \left( \sqrt{ r {m_2\over m_3}}\right)_I\eqno(III.28b)
$$
$$
s^c_{I23} = \left( \sqrt{ {1\over r} {m_2\over m_3}}\right)_I \eqno(III.28c)
$$
where $(m_{1,2,3})_I$ are the fermion mass eigenvalues of generations (1,2,3),
renormalized at the flavor scale $M_f$. 

Further aspects of this $U(2)$ theory of flavor can be found in \cite{BDH},
on which this section was based.

\vskip 9pt

{\bf III.7 The suppression of baryon and lepton number violation.}

\vskip 9pt

The standard model, for all its shortcomings, does provide an understanding 
for the absence of baryon and lepton number violation:
the field content simply does not allow any renormalizable interactions which
violate these symmetries.
This is no longer true when the field content is extended to become
supersymmetric; squark and slepton exchange mediate baryon and lepton number 
violation at
unacceptable rates, unless an extra symmetry, such as $R$ parity, is imposed 
on the theory.
It is worth stressing that some new symmetry, which in general we label by $X$,
really is required: the known gauge and spacetime symmetries are insufficient. 
The need for $X$  was first realised in the
context of a supersymmetric $SU(5)$ grand unified theory \cite{WI}.
As will become clear, there are a wide variety of possibilities 
for the $X$ symmetry. Matter parity \cite{DG}, $Z_N$ 
symmetries other than matter parity \cite{HS,BHR} and baryon or lepton 
numbers \cite{DH1} provide well known examples; 
each giving  a distinctive phenomenology.
One of the most fundamental questions in constructing supersymmetric models is
\cite{W}
{\it What is the origin of this extra symmetry needed to suppress baryon and 
lepton number violating processes?}

The $X$ symmetry must have its origin in one of the three categories of
symmetries which occur in field theory models of particle physics:
spacetime symmetries, gauge (or vertical) symmetries and flavor (or horizontal)
symmetries.
The $X$ symmetry is most frequently referred to as $R$ 
parity \footnote{$R_p$ was first introduced in a completely 
different context \cite{FF}.}, $R_p$, which is a $Z_2$
parity acting on the anti-commuting coordinate of superspace:
$\theta \to - \theta$.
We view this as unfortunate, since it suggests that the reason for the 
suppression of baryon and lepton number violation is to be found in spacetime 
symmetries, which certainly need not be the case.
$R_p$ can be viewed as a superspace analogue of the familiar discrete spacetime
symmetries, such as $P$ and $CP$.
In the case of $P$ and $CP$ we know that they can appear as accidental symmetries in gauge models which are sufficiently simple.
For example $P$ is an accidental symmetry of QED and QCD, while CP
is an accidental symmetry of the two generation standard model.
Nevertheless, in the real world $P$ and $CP$ are broken.
This suggests to us that discrete spacetime symmetries are not fundamental 
and should not be imposed on a theory, so that
if $R_p$ is a good symmetry, it should be understood as being an 
accidental symmetry resulting from some other symmetry. These
arguments can also be applied to alternative spacetime origins for $X$, 
such as a $Z_4$ symmetry on the coordinate $\theta$ \cite{HS}.
\footnote{
Clearly these arguments need not be correct: for example, it could be that both
$P$ and $CP$ are fundamental symmetries, but they have both been spontaneously 
broken.
However, in this case the analogy would suggest that $R_p$ is also likely to 
be spontaneously broken.}
Hence, while the symmetry $X$ could have a spacetime origin, 
we find it more plausible that it arises from gauge or flavor symmetries.

In this case what should we make of $R_p$? If it is a symmetry at all, it would
be an accidental symmetry, either exact or approximate. If $R_p$ is broken by
operators of dimension 3, 4 or 5, then a weak-scale, lightest superpartner (LSP)
would not be the astrophysical dark matter. The form of the $R_p$ breaking
interactions will determine whether the LSP will decay in particle detectors or
whether it will escape leaving a missing energy signature. The realization that
$X$ may well have an origin in gauge or flavor symmetries, has decoupled the two
issues of the suppression of $B$ and $L$ violation, due to $X$, and the
lifetime of the LSP, governed by $R_p$ \cite{BHR,H}.

At first sight, the most appealing origin for $X$ is an extension of the
standard model gauge group, either at the weak scale \cite{W}, or at the grand
unified scale \cite{SY}. An interesting example is provided by the crucial 
observation that adding
$U(1)_{B-L}$ \cite{SY}, or equivalently $U(1)_{T_{3R}}$, is sufficient to remove 
all renormalizable $B$ and $L$ violation from the low energy theory. 
Matter parity is a discrete subgroup of $U(1)_{B-L} \times U(1)_{T_{3R}}$. 
This is clearly seen in SO(10) \cite{CHH}, where the requirement that all
interactions have an even number of spinor representations immediately leads to
matter parity, generated by the $Z_2$ element
$$
X(SO(10)) = e^{i \pi (2T_{3_L} + 2 T_{3_R})} = e^{i \pi (N_{16} + N_{144} ...)}
\eqno(III.29)
$$
where $N_{16, 144, ...}$ is 1 for a 16, 144, ... representation.

However, this example has a gauge group with  rank larger than that of the
standard model, and the simplest way to spontaneously reduce the rank, for
example via the vev of a spinor 16-plet in $SO(10)$, leads to a large
spontaneous breaking of the discrete matter parity subgroup of $SO(10)$ 
\cite{M,KM}. Thus theories based on $SO(10)$ need a further ingredient to
ensure sufficient suppression of B and L violation of the low energy theory.
One possibility is that the spinor vev does not introduce the dangerous
couplings, which typically requires a discrete symmetry beyond $SO(10)$.
Alternatively the rank may be broken by a 
larger Higgs multiplets \cite{M}, for example the 126  
representation of $SO(10)$. Finally, if the reduction of rank occurs at low
energies, the resulting $R_p$ violating phenomenology may be acceptable 
\cite{KM}, however, the weak mixing angle prediction is then lost.
The flipped $SU(5)$ gauge group allows for models with renormalizable $L$
violation, but highly suppressed $B$ violation \cite{BrH}; however, these
theories also lose the weak mixing angle prediction.

There are other possibilities for $X$ to be a discrete subgroup of 
an enlarged gauge symmetry. Several $Z_N$
examples from $E_6$ are possible \cite{BHR}. Such a symmetry will be an anomaly
free discrete gauge symmetry, and it has been argued that if $X$ is discrete
it should be anomaly free
in order not to be violated by Planck scale physics \cite{KW}. With
the minimal low energy field content, there are only two such possibilities
which commute with flavor: the familiar case of matter parity, and a $Z_3$
baryon parity \cite{IR3}, which also prohibits baryon number violation from
dimension 5 operators. While the gauge origin of $X$ remains a likely
possibility, we are not aware of explicit compelling models which achieve this.

Finally we discuss the possibility that the $X$ symmetry is a flavor symmetry:
the symmetry which is ultimately responsible for the small parameters of the
quark and lepton mass matrices, and also of the squark and slepton mass
matrices, might provide sufficient suppression for $B$ and $L$ violation.
Indeed, this is an extremely plausible solution for the suppression of $L$
violation since the experimental constraints on the coefficients of the $L$ 
violating interactions are quite weak, and would be satisfied by having
amplitudes suppressed by powers of small lepton masses. 
However, the experimental constraints involving $B$ violation are so strong,
that suppression by small quark mass factors are insufficient
\cite{HK}. Hence the real challenge for these theories is to understand the
suppression of $B$ violation.

Some of the earliest models involving matter parity violation had a discrete 
spacetime \cite{HS} or gauge \cite{BrH} origin for $B$ conservation, but had
$L$ violation at a rate governed by the small fermion masses. This distinction
between $B$ and $L$ arises because left-handed leptons and Higgs doublets are
not distinguished by the standard model gauge group, whereas quarks are clearly
distinguished by their color. This provides a considerable motivation to search
for supersymmetric theories with matter parity broken only by the $L$ 
violating interactions.

It is not difficult to understand how flavor symmetries could lead to
exact matter parity. Consider a supersymmetric theory, with minimal field
content and gauge group, which has the flavor group $U(3)^5$ broken only by
parameters which transform like the usual three Yukawa coupling matrices.
The Yukawa couplings and soft interactions of the most general such effective
theory can be written as a power series in these breaking parameters, leading
to a theory known as weak scale effective supersymmetry \cite{HR}. The flavor
group and transformation properties of the breaking parameters are sufficient
to forbid matter parity violating interactions to all orders: each breaking
parameter has an even number of $U(3)$ tensor indices, guaranteeing that all
interactions must have an even number of matter fields. \footnote{ This point
was missed in \cite{HR} where $R_p$ was imposed unnecessarily as an additional
assumption. We believe that the automatic conservation of $R_p$ makes
this scheme an even more attractive framework as a model independent low energy
effective theory of supersymmetry.} To construct an explicit model along these
lines it is perhaps simplest to start with a $U(3)$ flavor group, with all
quarks and leptons transforming as triplets, but Higgs doublets as trivial
singlets. The $X$ symmetry is generated by the $Z_2$ element
$$
X(U(3)) =  e^{i \pi N_T} \eqno(III.30)
$$
where $N_T$ is the triality of the representation.
An exact matter parity will result
if the spontaneous breaking of this flavor group occurs only via fields with an
even triality.

\medskip

\noindent {\bf III.8 Conclusions.}

\medskip

The use of flavor symmetries to study both the fermion and scalar masses leads
to a new viewpoint. While fermion mass hierarchies remain a very fundamental
puzzle, the flavor-changing constraints are definitely {\em not} a problem for
supersymmetry; rather they are an advantage. Instead of a flavor-changing
problem, we have a tool that allows us to identify which flavor symmetries are
acceptable. Furthermore, many acceptable flavor symmetries lead to
flavor-changing phenomena beyond the standard model which should be discovered
in the not too distant future. Such discoveries provide the best hope for
progress on the fermion mass puzzle.

In this chapter I have pursued the idea that both fermion and scalar masses
should be constrained by the same approximate flavor symmetries. However,
fermion masses are supersymmetric while the soft scalar masses are not, so
that some decoupling of their symmetry behaviour is possible. Suppose that
fermion masses are understood in terms of physics at some flavor scale $M_f$.
If $M_f < M_{mess}$, the messenger scale of supersymmetry breaking discussed in
section I.5, then both fermion and scalar masses are subject to the same flavor
symmetries. However, if $M_{mess} < M_f$, as in models with low energy gauge
mediation of supersymmetry breaking \cite{ACW}, the soft operators can be
protected from the physics of fermion mass generation, leading to flavor
changing effects which are milder than those dictated by approximate flavor
symmetries. 

Broken flavor symmetries are the natural way to describe flavor sectors of
supersymmetric theories. For this reason the MSSM with universal boundary
conditions is badly flawed. We advocate replacing the universal boundary
condition of (III.15) with the modified boundary condition (III.16) which
results from the minimal necessary breaking of $G_{max} = U(3)^5$ \cite{HR}. 
Any relations between
$A_{U,D,E}$ or between $m^2_a$ should be viewed as probes of gauge unification
in the vertical direction. In general, corrections to (III.16) are
expected, as shown in (III.13d). Finally, in the simplest schemes, the Higgs
doublets are not related by flavor symmetries to the three generations of
matter, so the Higgs mass parameters should be taken to be independent of
$m^2_a$.

\noindent {\bf III.9 Appendix A}

\medskip

As an example of the $U(3)$ transformation conventions used in this chapter, 
I consider the first interaction of eq. III.12.
Making the transposition explicit, this is
$$
W = a\; q^T \mbox{\boldmath$\epsilon$}_U u \ h_2.\eqno(A1)
$$
Under $U(3)_q$ I take 
$$
q \to {\bf{L}}^* q\eqno(A2)
$$
Under $U(3)_u$ I take
$$
u \to {\bf{R}} u.\eqno(A3)
$$
Hence if I assign the transformation property
$$
\mbox{\boldmath$\epsilon$}_U \to {\bf{L}} \mbox{\boldmath$\epsilon$}_U 
{\bf{R}}^\dagger\eqno(A4)
$$
(A1) transforms to $q^T{\bf{L}}^\dagger{\bf{L}} \; \mbox{\boldmath$\epsilon$}_U
 \; {\bf{R}}^\dagger {\bf{R}}u \; h_2$,
and is therefore invariant.
I say that $\mbox{\boldmath$\epsilon$}_U$ transforms as $(3, \overline{3})$ 
under $(U(3)_{q}, U(3)_{u})$.

I write the scalar masses as
$$
V = q^T {\bf{m}}^2_q q^* + u^\dagger {\bf m}^2_u u\eqno(A5)
$$
so that ${\bf{m}}^2_q \to L{\bf{m}}^2_qL^\dagger, {\bf{m}}^2_u 
\to R{\bf{m}}^2_uR^\dagger $.
In building invariant terms it is useful to notice that 
$\mbox{\boldmath$\epsilon$}_U \mbox{\boldmath$\epsilon$}^\dagger_U, 
\mbox{\boldmath$\epsilon$}_D \mbox{\boldmath$\epsilon$}^\dagger_D$
 transform 
like ${\bf{m}}^2_q$, while 
$\mbox{\boldmath$\epsilon$}^\dagger_U \mbox{\boldmath$\epsilon$}_U$ transforms 
like ${\bf{m}}^2_u$.

\vskip 1in

{\large \bf IV. Supersymmetric Grand Unification} 

\vskip .2in

\noindent {\bf IV.1 Introduction}

How will we ever be convinced that grand unification, or string theory, or some
other physics at very high energies, is correct?
Two ways in which this could happen are:

\begin{enumerate}
\item The structure of the theory is itself so compelling and tightly 
constrained,
   and the links to observed particle interactions are sufficiently strong, that
   the theory is convincing and is accepted as  the standard viewpoint.
String theory is a candidate for such a theory, but connections to known
physics will require much further  understanding of the breaking of its 
many symmetries.
\item The theory predicts new physics beyond the standard model, which is
   discovered.
If the structure of the theory is not very tightly constrained, several such
predictions will be necessary for it to become convincing.
Grand unification is a candidate for such a theory, but as yet there have been
no discoveries beyond the standard model.
Supersymmetric grand unified theories do have a constrained gauge
structure, and this has led to the successful prediction of the weak mixing
angle at the percent level of accuracy \cite{GQW,DRW,DG,IR}.
\footnote{While giving the lectures at SLAC a bright spark in the audience
asked why I chose to quote $\sin^2 \theta = 0.231 \pm 0.003$, which suggests a
significance of 1\%, rather than using the well measured weak mixing angle as
input and quoting a prediction for the less well measured strong coupling
$\alpha_s = 0.126 \pm 0.013$, which looks to only have a significance of 10\%.
This is an excellent question. The reason I believe that the significance is
1\% rather than 10\% is as follows. 
Consider the $\sin^2 \theta / \alpha_s$ plane, with $\sin^2 \theta$ varying
from 0 to 1, and $\alpha_s$ varying from 0 to some large value $\alpha_s^c$
which is still perturbative. The area of this plane is $\alpha_s^c$, and it
could have been that the parameters lie anywhere in this plane.
The condition that the three gauge
couplings unify can be represented as a band in this
plane, with the width of the band representing the theoretical uncertainties,
such as the various threshold corrections. By sketching the plane, you can
convince yourself that the area of this band is given by $\alpha_s^c \Delta$,
where $\Delta$ is the theoretical uncertainty in $\sin^2 \theta$. Hence the
fraction of the area of the plane which the theory allows is $\Delta$, which is
of order 1\%,
and this is a measure of the significance of the prediction. This argument can
be rephrased by starting in some other basis for the parameters, 
\eg the space of 
$g_1, g_2$ and $g_3$ with $\alpha$ held fixed, but the conclusion will be the
same.}
While significant, this is hardly convincing.
Nevertheless, supersymmetric grand unified theories offer the prospect of many
further tests.
In this talk I make the case that experiments of this decade, and the next, 
allow
for the possibility that we might become convinced that grand unification is
correct.

\end{enumerate}

Any grand unified theory must have at least two sectors: the gauge sector, which
contains the gauge interactions, and the flavor sector containing the 
interactions which generate the quark and lepton masses.
In supersymmetric versions there are also the supersymmetry breaking
interactions.
I include the gaugino masses in the gauge sector, and the supersymmetry breaking
squark, slepton and Higgs masses and interactions in the flavor sector.
There are no known direct observable consequences of the interactions of the
superheavy gauge bosons: they are predicted to be too heavy even to mediate
proton decay at an observable rate.

I know of only one prediction in the gauge sector, other than $\sin^2 \theta$: 
ratios of the gaugino mass
parameters, $M_i, i=1,2,3$ for $U(1), SU(2)$ and $SU(3)$.
If the supersymmetry breaking is hard up to scales above the unification mass,
$M_G$, and if the breaking of supersymmetry in the gauge kinetic function is
dominantly $SU(5)$ preserving,
then $M_i$ will be independent of $i$ at $M_G$.
Beneath $M_G$, renormalizations induce splittings between the $M_i$, in fact
they scale exactly like the gauge couplings:
$M_i = \alpha_i M$.
The prediction of two gaugino mass ratios is a very important consequence of
super unification.
These predictions  occur in the gauge sector;
however, unlike the weak mixing angle, these predictions involve the
supersymmetry breaking sector, and even if the supersymmetry breaking is hard at
$M_G$, there are situations when they are broken \cite{HMG}.
Furthermore, these relations can occur without grand unification.
\footnote{Suppose supersymmetry is broken in a sector
which communicates with the observable sector only via standard model gauge
interactions.
Then one expects $M_i \propto \alpha_i$ as before.
The constant of proportionality is not guaranteed to be
independent of $i$, although such an independence follows if the particles
communicating the supersymmetry breaking fill out complete $SU(5)$ multiplets,
as suggested by the weak mixing angle prediction.}

\noindent {\bf IV.2 Flavor Signals Compared}

Fortunately, the flavor sector has many signatures,
listed in Table 3 in 5 categories.
Proton decay \cite{PS,GG} and neutrino masses \cite{GRS,Y} are the earliest
and most well-known signatures of grand unification.
However, the theoretical expectation for these classic signals is plagued by
a power dependence on an unknown superheavy mass scale.
For neutrino masses this is the right-handed Majorana mass $M_R$.
If we naively set 
$m_{\nu_i} = m^2_{u_i}/M_R$ with $M_R = M_G = 2 \times 10^{16}$ GeV, 
then all three neutrino masses are too small to be detected in any laboratory 
experiment, although they could lead to MSW oscillations in the sun.

While the many hints for detection of neutrino oscillations are extremely 
interesting,
and theorists are full of ideas for suppressing $M_R$, if we fail to detect 
neutrino masses then we learn very little about grand unification.
On  the other hand, several observations hint at the presence of neutrino 
masses, and measurements of neutrino mass ratios and 
mixing angles would provide a very important probe of the flavor structure of 
unified models.

\newpage

\centerline {\bf Table 3}

\begin{center}
\vskip 20pt
\begin{tabular}{|l|c|c|c|}
\hline
&Requires&``Present''&Requires\cr
&BSM&in all$^\dagger$&Susy breaking\cr
&discovery&models&hard at $M_G$\cr
\hline
(I) \ $p$ \ decay& $\surd$&No&No\cr
&&&\cr
\hline
(II) \ $\nu$ masses&$\surd$&No&No\cr
&&&\cr
\hline
(III) \ $u,d,e$&No&No&No\cr
masses and mixings&&&\cr
\hline
(IV)\ $\tilde{u}, \tilde{d},\tilde{e}$&$\surd$&$\surd$&$\surd$\cr
masses&&&\cr
\hline
(V)\  $L_{e,\mu,\tau}$ and &$\surd$&$\surd$&$\surd$\cr
CP violation&&&\cr
\hline
\end{tabular}
\end{center}

\begin{quotation} Characteristic features of the 5 flavor tests of 
supersymmetric grand unification. 
\end{quotation}

The leading supersymmetric contribution to the proton decay rate is 
proportional to $M_H^{-2}$, 
\cite{W,SY} where $M_H$ is a model dependent parameter, which arises from the 
unified symmetry breaking sector of the theory.
The simple expectation that $M_H\simeq M_G$ is excluded as it produces too short
a proton lifetime \cite{W,SY}.
There are many mechanisms that effectively allow $M_H$ to be enhanced, 
thereby stabilizing
the proton, but there is no argument, which I would defend, demonstrating that 
proton decay will be within reach of future experiments.
If we are lucky, proton decay may be discovered, and the decay modes
and branching ratios will probe flavor physics in an important way.
However, as for neutrino masses, if a signal is not
seen, little of use is learnt about the question of grand unification, hence
the ``No" in the middle column of Table 3.

The third signature of the flavor sector of grand unified theories is provided 
by relations amongst the masses and mixings of the quarks and 
charged leptons, which was also first studied in the 1970s \cite{CEG}.
This signature has the very great advantage over all others that data exists:
there is no need for discoveries beyond the standard model.
Since the late 70s this field has developed considerably, in step with our 
continually increasing knowledge of the quark and lepton masses and the 
Kobayashi-Maskawa matrix elements. These signatures are based on the hope that 
the flavor interactions which generate the 
fermion masses are relatively simple, involving few enough parameters that 
relations among the 13 observables can be derived.
While there is no guarantee that this is true, it is an assumption which is 
reasonable and which could have an enormous payoff.
A considerable fraction of high energy physics experiments aim at extracting
more precise valves for the quark masses and mixings; each time an
error bar is reduced, this probe of grand unification becomes
more incisive.
Among the interesting results obtained so far are:

\begin{quotation}
$\bullet$        Evolution of the $b$ and $\tau$ Yukawa couplings to high 
energies in the standard model does not lead to their unification, 
as expected from the simple $SU(5)$ boundary condition.
Such a unification does work well if evolution is done with weak scale 
supersymmetry and a heavy top quark \cite{B1,B2,B3,B4}.
\end{quotation}
\begin{quotation}
$\bullet$ The unification of the three Yukawa couplings of the heavy 
generation in 
the MSSM \cite{ALS}, expected from a simple $SO(10)$ boundary condition, can 
occur perturbatively only if 165 GeV $< m_t <$ 190 GeV. \cite{HRS}.
\end{quotation}
\begin{quotation}
$\bullet$ It is possible to construct $SO(10)$ models where all observed 
fermion masses and mixings are generated from just 4 interactions.
Seven of the 13 flavor parameters are predicted \cite{ADHRS}.
\end{quotation}
\begin{quotation}
$\bullet$ The observed quark masses and mixings may be consistent with several 
patterns 
of the Yukawa matrices at the unification scale in which many of the entries 
are zero, suggesting they have a simple origin \cite{RRR}.
\end{quotation}

I have discussed the first three signatures of Table 1, stressing that only for 
fermion mass relations do we have any useful data, and stressing that
none of these signatures is a necessary consequence of grand unification.
These features are shown in the first two columns of the Table.
We must now discuss supersymmetry breaking, which is relevant for the third 
column of Table 3.
The fundamental origin of the first three signatures (baryon number violation,
lepton number violation, and Yukawa coupling
relations) does not depend on supersymmetry breaking.
However, for the last two signatures, the supersymmetry
breaking interactions of the low energy effective theory contain all the
information relevant to the signals.

A crucial question for these two signatures is: at what
scale do the interactions which break supersymmetry become soft?
This has nothing to do with the size of the parameters which violate 
supersymmetry -- they are of order the weak scale.
At any energy scale, $\mu$, we can consider our theory to be a local effective 
field theory.
What is the ``messenger scale", $M_{mess}$, above which the supersymmetry 
breaking parameters, such as squark 
and gluino masses, do not arise from a single local interaction?
Consider models where supersymmetry is broken spontaneously in a sector 
with a single mass scale, $M$, and is communicated to the observable sector 
by the known gauge interactions \cite{ACW,DF}.
It is only when the particles of mass $M$ are integrated out of the theory that
local interactions are generated for squark and gluino masses.
Hence for these models the messenger scale is given by $M_{mess} = M$, 
which is of order $M_W/\alpha$, or 10 TeV.

The breaking of supersymmetry in a hidden sector of $N=1$ supergravity 
theories \cite{BFS,HLW} has become a popular view
(although it is not satisfactory in several respects).
The interactions which generate squark and slepton masses are produced 
when supergravity auxiliary fields are eliminated from the theory, and hence 
are local at all energies up to the Planck scale, giving 
a messenger scale $M_{mess} = M_{Pl}$.
For signatures IV and V the critical question is whether $M_{mess}$ is larger 
or smaller than $M_G$, the unification mass.
If $M_{mess} \ll M_G$ then the local interactions which break supersymmetry
are produced at energies beneath $M_G$,
and hence these interactions are not renormalized by the interactions of 
the unified theory.
On the other hand, if $M_{mess} \geq M_G$, then the supersymmetry
breaking interactions appear as local interactions in the grand unified
theory itself. At energies above $M_G$ they take a form which is constrained by
the unified symmetry. Furthermore, they are modified by radiative corrections
induced by the unified theory, giving low energy signals which are not power
suppressed by $M_G$ \cite{HKR}.

For example, in any grand unified theory in which $\tilde{u}, \tilde{u}^c$ and
$\tilde{e}^c$ are unified in the same irreducible representation, the 
unified theory
will possess $m^2_{\tilde{u}} = m^2_{\tilde{u}^c} = m^2_{\tilde{e}^c}$.
When the unified gauge symmetry is broken, such relations can be modified 
both radiatively and at tree level.
However, it has been shown that in all models where the weak
mixing angle is a significant prediction of the theory,
there will be two scalar superpartner mass relations for each of the 
lightest generations \cite{CH}.

It is possible that the gauge forces are unified but the low energy matter 
particles are not, for example  $\tilde{u}, \tilde{u}^c$ and
$\tilde{e}^c$ could lie in different irreducible representations of the unified
group. In this case the unified gauge group clearly does not lead to scalar
mass relations amongst the light states. While this situation is a logical
possibility, I do not find it very plausible. It is not straightforward to
construct such theories and maintain an understanding for the smallness of the
flavor mixing angles of the Kobayashi-Maskawa mixing matrix. Much more likely
is the possibility that the light mass eigenstate fields  $\tilde{u}, 
\tilde{u}^c$ and
$\tilde{e}^c$ lie dominantly in one irreducible representation, but have small
components in other representations \cite{DP}.
This happens automatically in
Froggatt-Nielsen theories of fermion masses \cite{FN} which rely heavily on 
mass mixing between heavy and light states.
Such small mixings will lead to
corresponding small deviations from the exact unified scalar mass relations
of \cite{CH}. In principle these shifts in the scalar mass eigenvalues 
would allow sparticle spectroscopy to be used as a probe of the unified theory
\cite{DP}. However, I doubt they will be big enough to be directly seen in
spectroscopy. This is because the mass
mixings also induce flavor changing effects in the scalar sector, and these are
powerfully constrained by experiment. Since this phenomenon occurs at tree
level, it is likely to dominate over the flavor changing effects that the 
unified theory will induce at the loop level \cite{HKR}, and hence will become
one of the most important constraints on building theories of fermion masses
using the Froggatt-Nielsen method. Hence, I think that simple scalar mass 
relations are likely to result in unified theories, while
the flavor changing phenomenology will probe details of the flavor structure 
of the unified theory.

\noindent {\bf IV.3 Flavor Changing and CP Violating Signals}

Riccardo Barbieri and I have recently shown that a new class of signatures 
arises in supersymmetric theories which unify the top 
quark and $\tau$ lepton, and which have a high messenger scale  
$M_{mess} > M_G$ \cite{BH}. These effects are induced by
radiative corrections involving the large
top Yukawa coupling of the unified theory, $\lambda_{tG}$.
The most promising discovery signatures are lepton flavor violation, 
such as $\mu\to e\gamma$\cite{BH,BHS} and electric dipole 
moments for the electron and neutron, $d_e$ and $d_n$ \cite{DH,BHS}.

These signatures are complementary to the classic tests of proton decay and 
neutrino masses,
as shown in the last two columns of Table 1.
We believe that these new signatures are much less model dependent than 
the classic tests: they are present in a very wide range of models with 
$M_{mess} > M_G$. A second crucial point, when comparing with the classic tests,
is the size of these signals, which does not depend on the power of an 
unknown superheavy mass.

A complete calculation in the minimal $SU(5)$ and $SO(10)$ models \cite{BHS} 
concludes that searches for the $L_i$ and CP violating signatures provide the 
most
powerful known probes of supersymmetric quark-lepton unification with 
supersymmetry breaking generated at the Planck scale.
For example, an experiment with a sensitivity of 10$^{-13}$ to B.R.
$(\mu\to e\gamma)$
would probe (apart from a small region of parameter space where cancellations in
the amplitude occur) the $SU(5)$ model to $\lambda_{tG} =1.4$ and 
$m_{\tilde{e}_R}=$ 
100 GeV, and would explore a significant portion of parameters space for 
$m_{\tilde{e}_R}=$ 300 GeV.
In the $SO(10)$ case, where the present bound on $\mu\to e\gamma$ is already 
more stringent than the limits from high energy accelerator experiments, a 
sensitivity of 10$^{-13}$ would probe the theory to $\lambda_{tG} =$ 1.25 and 
$m_{\tilde{e}_R}$ close to 1 TeV.

Which search probes the theory more powerfully: rare muon processes or the
electric dipole moments? In the minimal SU(5) theory, the electric dipole
moments are very small so that the rare muon processes win. In the minimal
SO(10) theory,
the electric dipole moments are proportional to $\sin
\phi$ where $\phi = \phi_d - 2 \beta$, where $- \beta$ is the phase of the
Kobayashi-Maskawa matrix element $V_{td}$, and $\phi_d$ is a new phase.
There is a simple relation between B.R. $(\mu \rightarrow e \gamma)$ and $d_e$:
$$ 
{\abs{d_e} \over 10^{-27} \mbox{e cm}} = 1.3 \; \sin \phi
\sqrt{ { \mbox{B.R.} (\mu \rightarrow e \gamma) \over 10^{-12}}}. \eqno(IV.1)
$$
For $\sin \phi = 0.5$, the present limits imply that the
processes have equal power to probe the theory. The analysis of the data from
the MEGA experiment should put the rare muon decay ahead, but
eventually $d_e$ may win because it falls only as the square of the
superpartner mass, whereas the rare muon decay rate falls as the fourth power.
At some point these processes could force the selectron masses to be higher
than is reasonable from the viewpoint of electroweak symmetry breaking,
discussed in section II.3.

Similar new flavor-changing tests of supersymmetric quark-lepton unification 
occur in the hadronic sector, where the best probes are non-standard model 
contributions
to $\epsilon, b\to s\gamma$ and to CP violation in neutral $B$ meson decays 
\cite{BHS1}.
These signals could provide a powerful probe of the flavor sector of unified
theories.
However, unlike the lepton flavor violating and electric dipole signatures, 
they must
be distinguished from the standard model contribution, and they are small 
when the gluino is heavy due to a gluino focussing effect on the squark masses.

Unified flavor sectors which are more complicated than the
minimal ones lead to a larger range of predictions for these signals.
There may be additional sources of flavor and CP violation other than those 
generated by the top Yukawa coupling.
While cancelling contributions cannot be ruled out, they are unlikely to lead 
to large suppressions.
Many other sources could provide effects which are larger than those generated 
by $\lambda_{tG}$, and hence it is reasonable to take the top contribution as 
an indication of the minimum signal to be expected.

\vskip .2in

\noindent {\bf IV.4 The top quark origin of new flavor and CP violation}

\medskip

At first sight, it is surprising that the top quark Yukawa coupling should lead
to any violation of $L_e$ or $L_\mu$. What is the physical origin of this
effect, and why is it not suppressed by inverse powers of $M_G$? The answer
lies in new flavor mixing matrices, which are analogous to the
Kobayashi-Maskawa matrix.

In the standard model the quark mass eigenstate basis is reached by making
independent rotations on the left-handed up and down type quarks, $u_L$ and
$d_L$. However, these states are unified into a doublet of the weak SU(2) gauge
group: $Q = (u_L, d_L)$. A relative rotation between $u_L$ and $d_L$ therefore
leads to flavor mixing at the charged W gauge vertex. This is the well-known
Cabibbo-Kobayashi-Maskawa mixing. With massless neutrinos, the standard model
has no analogous flavor mixing amongst the leptons: the charged lepton mass
eigenstate basis can be reached by a rotation of the entire lepton doublet $L =
(\nu_L, e_L)$.

How are these considerations of flavor mixing altered in supersymmetric unified
theories? There are two new crucial ingredients. The first is provided by
weak-scale supersymmetry, which implies that the quarks and leptons have scalar 
partners. The mass eigenstate basis for these squarks and sleptons requires
additional flavor rotations. As an example, consider softly broken
supersymmetric QED with three generations of charged leptons. There are three
arbitrary mass matrices, one for the charged leptons, $e_{L,R}$, and one each
for the left-handed and right-handed sleptons, $\tilde{e}_L$ and $\tilde{e}_R$.
To reach the mass basis therefore requires a relative rotation between 
$e_{L,R}$ and $\tilde{e}_{L,R}$, resulting in a flavor mixing matrix at the
photino gauge vertex. These matrices were called ${\bf W}^{e_L,e_R}$ in section
III.3.

In supersymmetric extensions of the standard model, these additional
flavor-changing effects are known to be problematic. With a mixing angle
comparable to the Cabibbo angle, a branching ratio for  $\mu \rightarrow 
e \gamma$ of order $10^{-4}$ results. In the majority of 
supersymmetric models which have been constructed, such flavor-changing
effects have been suppressed by assuming that the origin of supersymmetry
breaking is flavor blind. In this case the slepton mass matrix is proportional
to the unit matrix. The lepton mass matrix can then be diagonalized by
identical rotations on $e_{L,R}$ and $\tilde{e}_{L,R}$, without introducing
flavor violating mixing matrices at the gaugino vertices.
{\it Slepton degeneracy renders lepton flavor mixing matrices non-physical.}

The unification of quarks and leptons into larger multiplets provides the
second crucial new feature in the origin of flavor mixing.  The weak 
unification of
$u_L$ and $d_L$ into $q_L$ is extended in SU(5) to the unification of $q_L$ with
$u_R$ and $e_R$ into a 10 dimensional multiplet $T(q_L, u_R, e_R)$. Since higher
unification leads to fewer multiplets, there are fewer rotations which can be
made without generating flavor mixing matrices.

In any supersymmetric unified model there must be at least two coupling
matrices, $\mbox{\boldmath$\lambda$}_1$ and 
$\mbox{\boldmath$\lambda$}_2$, which describe quark masses.
If there is only one such matrix, it can always be diagonalized without
introducing quark mixing. One of these coupling matrices, which we take to be 
$\mbox{\boldmath$\lambda$}_1$, must contain the large coupling, 
$\lambda_t$, which is
responsible for the top quark mass. We choose to work in a basis in which 
$\mbox{\boldmath$\lambda$}_1$ is diagonal. 
The particles which interact via $\lambda_t$ are
those which lie in the same unified multiplet with $t_L$ and $t_R$. In all
unified models this includes a right-handed charged lepton, which we call
$e_{R_3}$. This cannot be identified as the mass eigenstate $\tau_R$, because
significant contributions to the charged lepton masses must come from the
matrix $\mbox{\boldmath$\lambda$}_2$, which is not diagonal.

The assumption that the supersymmetry breaking mechanism is flavor blind,
leads to mass matrices for both $\tilde{e}_L$ and $\tilde{e}_R$ which are
proportional to the unit matrix at the Planck scale, $M_{Pl}$. As we have seen,
without unified interactions lepton superfield rotations can diagonalize the
lepton mass matrix without introducing flavor mixing matrices. However, the
unification prevents such rotations: the leptons are in the same multiplets as
quarks, and the basis has already been chosen to diagonalize 
$\mbox{\boldmath$\lambda$}_1$.
As the theory is renormalization group scaled to lower energies, the
$\lambda_t$ interaction induces radiative corrections which suppress the mass
of $\tilde{e}_{R_3}$ beneath that of $\tilde{e}_{R_1}$ and $\tilde{e}_{R_2}$. 
Beneath $M_G$ the superheavy particles of the theory can be decoupled, leaving
only the interactions of the minimal supersymmetric standard model. Now that
the unified symmetry which relates quarks to leptons is broken, a lepton mass
basis can be chosen by rotating lepton fields relative to quark fields.
However, at these lower energies the sleptons are no longer degenerate, so that
these rotations do induce lepton flavor mixing angles.
{\it Radiative corrections induced by $\lambda_t$ lead to slepton
non-degeneracies, which render the lepton mixing angles physical.}

This discussion provides the essence of the physics mechanism for $L_{e,\mu,
\tau}$ violation in superunified models. It shows the effect to be generic to
the idea of quark-lepton unification, requiring only that supersymmetry survive
unbroken to the weak scale, and that supersymmetry breaking be present at the
Planck scale. The imprint of the unified interactions is made on the
soft supersymmetry breaking coefficients, including the scalar trilinears,
which are taken to be universal at the Planck scale. 
Eventually this imprint will be
seen directly by studying the superpartner spectrum, but it can also be probed
now by searching for $L_{e,\mu,\tau}$ and CP violating effects.

The above discussion assumed a universal scalar mass at high energies. We
argued in Chapter III that it is preferable to replace this ad hoc form with
scalar masses that are the most general allowed by an appropriate flavor group,
$G_f$. This group solves the ``1--2" flavor problem, as discussed in section 
III.3, but the ``1,2--3" flavor signature discussed here, which results from
the large splitting between the scalars of the third generation and those of
the lighter two generations, will persist.

\noindent {\bf IV.5 Summary}

Supersymmetric grand unified theories are a leading
candidate for physics beyond the standard model because

\begin{quotation}$\bullet$ They provide an elegant group theoretic
understanding of the gauge quantum numbers of a generation.
\end{quotation}

\begin{quotation}$\bullet$ $\sin^2 \theta$ is the only successful prediction 
of any parameter of the standard model at the percent level of accuracy.
\end{quotation}

I have not yet mentioned the most crucial experimental hurdle which these
theories must pass: superpartners must be discovered at the weak scale.
Without this, I will never be convinced that these theories are correct.
As I write, I imagine the sceptics who may read this (I dare to hope!) 
saying ``suppose 
by 2010 we have measured neutrino masses
and mixing angles, seen proton decay and other rare processes such as 
$\mu\to e\gamma$,
$d_e$ and $d_n$, found non-standard CP violation in $B$ meson decays, 
and that we have even discovered superpartners and measured their masses.
This still will not convince me that the theory behind this physics is 
quark-lepton unification.''
My reply is

\begin{quotation}$\bullet$ These discoveries will not necessarily make 
quark-lepton unification convincing, 
but they will make it the standard picture.\end{quotation}
\begin{quotation}$\bullet$ These discoveries might make a particular model of
quark-lepton unification completely convincing.
\end{quotation}

There is certainly no guarantee of the latter point, but let me illustrate 
it with an optimistic viewpoint.
There are millions of possible flavor sectors of unified models.
Some are so complicated that, if this is the way nature is, we are unlikely to 
ever uncover this structure from low energy experiments alone.
Others are very simple with few interactions and parameters.
Why should nature be kind to us and provide a simple flavor sector with few
interactions? Quite apart from our general belief that the underlying laws of
physics will be simple, I think that the answer is illustrated by the $U(2)$
model of section III.7. A flavor symmetry provides a convincing solution to the
flavor changing problem. Since it must severely constrain the scalar
sector, it is expected to also severely restrict the fermion mass operators. 
The most constrained scheme which  I know has 10 parameters (8 flavor and 2 
supersymmetry breaking) to describe all the flavor physics signals.
As an example, consider something in between with, say, 15 parameters (eg. 12 
flavor and 3 supersymmetry breaking).
This has two parameters more than the flavor sector of the standard model.
Suppose that we discover such a unified model with 
these two parameters correctly describing: the entire superpartner spectrum, the neutrino
masses and mixing angles and the magnitudes of the non-standard model signals for 
$\mu \to e\gamma, d_e, d_n$ and $ B$ meson CP violation,
and the masses of the two Higgs bosons, the pseudoscalar boson 
and the charged Higgs boson.
It is certainly an optimistic scenario, but it is one which I would find 
convincing.

\vskip .25in

\centerline{\bf V. The High Energy Frontier}

\medskip

What are the liveliest debates at the high energy frontier today?
Particle physics, like other branches of physics, is driven first and foremost
by experimental discoveries.
Many experimental discoveries laid the groundwork for the development of the
gauge structure of the standard model, and we will need many further
experiments to guide us beyond. Hence, it is not surprising that the dominant
debate of the field is about which accelerators should be built and which 
experiments should be done.

The phenomena uncovered by experiments have led to a stunning array of
theoretical developments over the last 30 years.
These theoretical tools allowed the construction of the standard model.
A dominant debate in theoretical circles is whether the tools of point
particle field theories and their symmetries will take us much further, or
whether further tools, such as string theory are necessary.

There is no doubt that there are limits to point particle gauge theory, the
clearest of which is that they cannot describe gravity.
Nevertheless, point particle gauge theories and their symmetries are an
extraordinarily rich and powerful tool.
In these lectures I have explored the possibility that they provide a deeper
understanding of many of the outstanding questions of particle physics.

$\bullet$ A dynamical origin of electroweak symmetry breaking as a heavy top 
quark effect.\\

$\bullet$ A flavor symmetry origin for the pattern of fermion masses and 
mixing.\\

$\bullet$ A unified gauge symmetry - allowing for a highly
constrained and predictive theory of flavor, in addition to the well known
picture of a unified family and unified gauge couplings. \\

It is extraordinary that such a comprehensive vision of particle interactions
has been developed.
It seems unlikely that a complete picture of particle physics can be 
constructed without non-perturbative dynamics entering at some point; but 
what is that point? 
It is possible that the failure to develop a comprehensive vision of particle 
physics beyond the standard model based on either technicolor or a
composite Higgs is because in these cases the issue of non-perturbative 
dynamics provides a barrier at the very first step.  
The vision developed here is largely perturbative and is based on weak-scale
supersymmetry, a heavy top quark leading to perturbative dynamics for 
electroweak symmetry breaking, and perturbative unification. 
The only new non-perturbative dynamics beneath the Planck scale occurs
in the supersymmetry breaking sector, which I have not discussed. Fortunately,
there are many experimentally testable aspects af the theory which follow from
a few minimal assumptions, and no detailed understanding, about how 
supersymmetry breaking occurs. 
Measurements of the superpartner masses will provide a crucial guide as
to how the supersymmetry breaking interactions should be generated.

The vision of weak scale supersymmetry and perturbative unification
receives considerable motivation from precision electroweak measurements, 
but only further
experiments will prove whether these ideas are correct.  The discovery of
supersymmetry at the weak scale would be a revolution for High Energy Physics,
as important as any the field has seen, heralding a new era. Decades of
experimentation would be needed to fully elucidate the ramifications of this
new symmetry; for example, measurements of the many new flavor observables 
would provide a new handle on the flavor problem.

{\bf Acknowledgements}

I am grateful to the organizers for arranging a very stimulating school.

\vskip 1in

\centerline{\bf Figure Caption}

\medskip

Upper bounds on superpartner and Higgs boson masses which follow 
from requiring a limit to the amount of fine tuning among parameters. 
This figure applies to
the supersymmetric extension of the standard model with minimal field content,
with all scalar masses taken equal at the unification scale, and similarly for 
the three gaugino masses. The upper extent
of the lines for each particle correspond to $\tilde{\gamma} = 10$, the error
bar symbol to  $\tilde{\gamma} = 5$, and the squares to the masses which
result from minimizing the amount of fine tuning. This figure was supplied to
me by Greg Anderson; for further figures see \cite{AC}.

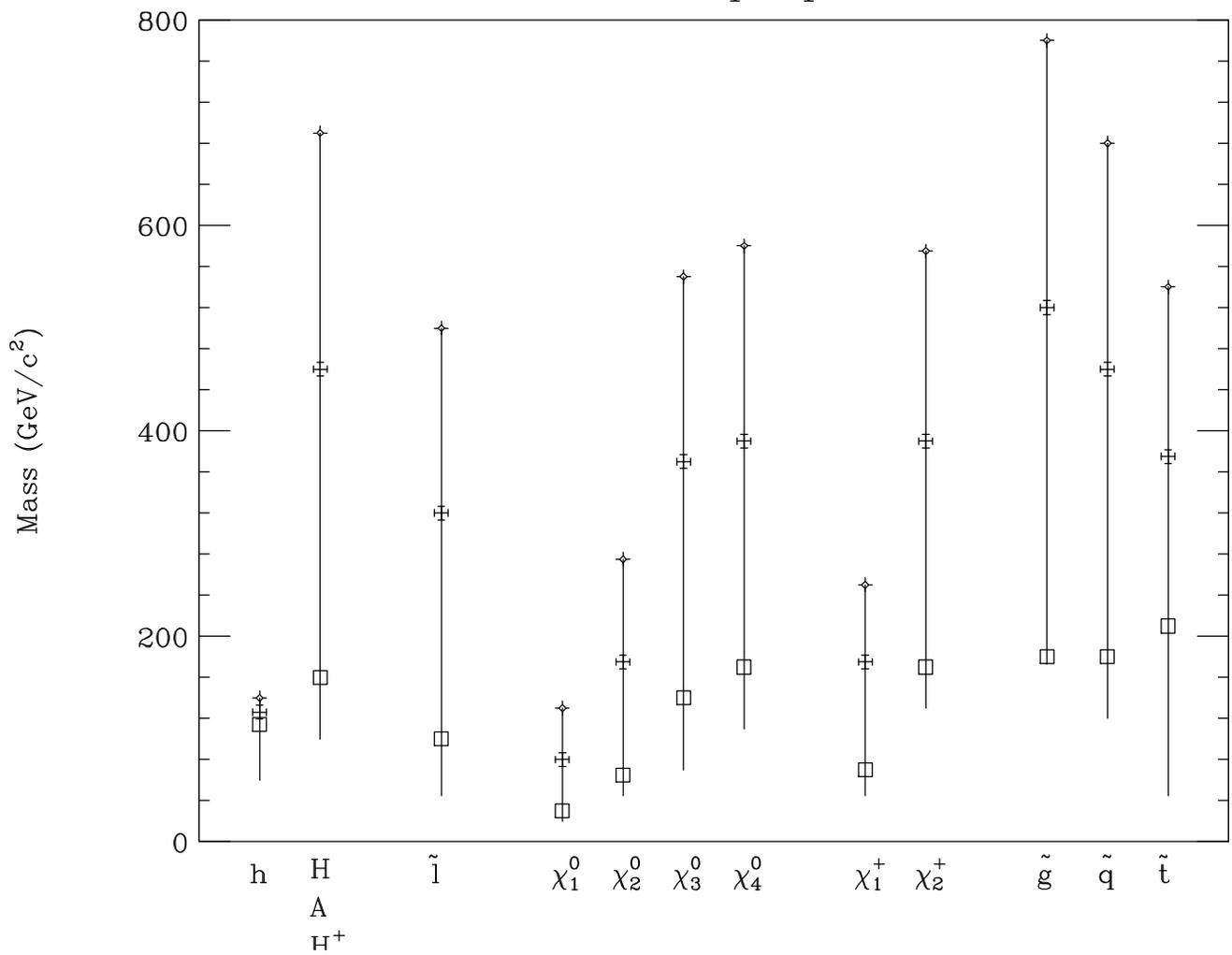
\end{document}